\documentstyle[preprint,eqsecnum,aps,epsf]{revtex}
\newif\iftightenlines\tightenlinesfalse
\tightenlines\tightenlinestrue
\def\lsim{\:\raisebox{-0.5ex}{$\stackrel{\textstyle<}{\sim}$}\:}
\def\gsim{\:\raisebox{-0.5ex}{$\stackrel{\textstyle>}{\sim}$}\:}
\def\ie{{\it i.e.}}
\def\half{{\textstyle{1 \over 2}}}
\def\quarter{{\textstyle{1 \over 4}}}
\def\mz{M_Z}

\def\mgut{M_{GUT}}
\def\mpl{M_{Planck}}

\def\eslt{\not\!\!{E_T}}
\def\to{\rightarrow}

\def\te{\tilde e}

\def\tu{\tilde u}
\def\tb{\tilde b}

\def\td{\tilde d}
\def\tQ{\tilde Q}
\def\td{\tilde d}

\def\tL{\tilde L}
\def\tH{\tilde H}
\def\tst{\tilde t}
\def\ttau{\tilde \tau}

\def\tg{\tilde g}
\def\tnu{\tilde\nu}
\def\tell{\tilde\ell}
\def\tq{\tilde q}
\def\tw{\widetilde W}
\def\tz{\widetilde Z}
\begin{document}
\draft
\preprint{\vbox{\baselineskip=14pt%
   \rightline{FSU-HEP-000222}\break 
   \rightline{UCCHEP/9-00}
   \rightline{UH-511-956-00}
}}

\title{IMPACT OF PHYSICAL PRINCIPLES \\ AT VERY HIGH ENERGY SCALES\\
ON THE SUPERPARTICLE MASS SPECTRUM}

\author{Howard Baer$^1$, Marco A. D\'\i az$^{1,2}$, Pamela Quintana$^1$
and Xerxes Tata$^3$}
\address{
$^1$Department of Physics,
Florida State University,
Tallahassee, FL 32306 USA
}
\address{
$^2$Departmento de F\'\i sica,
Universidad Cat\'olica de Chile,
Santiago 6904411, Chile
}
\address{
$^3$Department of Physics and Astronomy,
University of Hawaii,
Honolulu, HI 96822 USA
}
%
%\date{\today}
\maketitle
\begin{abstract}

We survey a variety of proposals for new physics at high scales that
serve to relate the multitude of soft supersymmetry breaking parameters
of the MSSM. We focus on models where the new physics results in
non-universal soft parameters, in sharp contrast with the usually
assumed mSUGRA framework. These include {\it i}) $SU(5)$ and $SO(10)$
grand unified (GUT) models, {\it ii}) the MSSM plus a right-handed
neutrino, {\it iii}) models with effective supersymmetry, {\it iv})
models with anomaly-mediated SUSY breaking and gaugino mediated
SUSY breaking, {\it v}) models with
non-universal soft terms due to string dynamics, and {\it vi}) models
based on $M$-theory.  We outline the physics behind these models, point
out some distinctive features of the weak scale sparticle spectrum, and
allude to implications for collider experiments. To facilitate future
studies, for each of these scenarios, we describe how collider events
can be generated using the program ISAJET. Our hope is that
detailed studies of a variety of alternatives will help point to the
physics underlying SUSY breaking and how this is mediated to the observable
sector, once sparticles are discovered and their properties measured.

\end{abstract}

\medskip


%%%%%%%%%%%%%%%%%% MAIN TEXT %%%%%%%%%%%%%%%%%%%%%%%%%%%%%%%%%%%%%%%%%%%%%%%
\section{Introduction}

The Minimal Supersymmetric Standard Model (MSSM) is a 
well-motivated extension of the Standard Model (SM) that includes
broken supersymmetry (SUSY) at the weak scale\cite{mssm}. 
To construct the MSSM, one
postulates:
\begin{itemize}
\item the gauge group and the matter content of the SM, 
where the various fields are replaced by superfields:
\end{itemize}
\begin{eqnarray*}
\hat{Q}_i=\left( \begin{array}{c} \hat{u}_i \\ \hat{d}_i\end{array}\right) ,\
\hat{L}_i=\left( \begin{array}{c} \hat{\nu}_i \\ \hat{e}_i\end{array}\right) ,\
\hat{U}^c_i,\ \hat{D}^c_i,\ \hat{E}^c_i ,
\end{eqnarray*}
where $i=1,2,3$ corresponding to the various generations;
\begin{itemize}
\item an extended Higgs sector that includes two different $SU(2)$
doublet Higgs superfields  
%which allow complete anomaly cancellation and give
%mass to both up and down-type quarks and leptons,
%\end{itemize}
\begin{eqnarray*}
\hat{H}_u({\bf 2})=\left( \begin{array}{c} \hat{h}_u^+ \\ \hat{h}_u^0
\end{array}\right) ,\
 {\rm and}\ \hat{H}_d({\bf \bar{2}})=
\left( \begin{array}{c} \hat{h}_d^- \\ \hat{h}_d^0
\end{array}\right) ;
\end{eqnarray*}

to allow superpotential Yukawa couplings (and
hence, masses) for
both up and down type fermions.
The introduction of two doublets of Higgsinos is also just right to cancel
the chiral anomaly from the gauginos.
\end{itemize}
\begin{itemize}
%\item a supersymmetric, renormalizable Lagrangian which is invariant under the
%$SU(3)_C\times SU(2)_L\times U(1)_Y$ gauge symmetry of the SM,
\item an $R$-parity conserving renormalizable
superpotential,~\footnote{Our sign convention for the $\mu$-term is 
defined by the chargino and neutralino mass matrices given in
Eqs.~(33) and (34) of the review by X.~Tata, Ref.~\cite{mssm}.}
\begin{eqnarray*}
 \hat{f}=\mu\hat{H}_u^a\hat{H}_{da}+f_u\epsilon_{ab}\hat{Q}^a\hat{H}_u^b
\hat{U}^c+f_d\hat{Q}^a\hat{H}_{da}\hat{D}^c+f_e\hat{L}^a\hat{H}_{da}\hat{E}^c
+\cdots,
\end{eqnarray*}
where $\epsilon_{ab}$ is the completely antisymmetric $SU(2)$ tensor
with $\epsilon_{12}=1$, and the ellipses refer to Yukawa couplings for
the second and third generations; 
\end{itemize}
\begin{itemize}
\item soft supersymmetry breaking (SSB) terms consistent with Lorentz
invariance and SM gauge invariance, 
\end{itemize}
\begin{eqnarray*}
{\cal L}_{soft}=-\sum_r m_r^2|\phi_r|^2&-&{1\over 2}\sum_\lambda
M_\lambda \bar{\lambda}_\alpha\lambda_\alpha +\left[ B\mu\tH_d\tH_u +
h.c \right ] \\ &+& \left[A_u
f_u\epsilon\tQ\tH_u\tu_R^\dagger +A_d f_d\tQ\tH_d\td_R^\dagger+A_e
f_e \tL\tH_d\te_R^\dagger +\cdots + h.c. \right ],
\end{eqnarray*}
where contraction over the $SU(2)$ indices is understood, and the
ellipses again refer to terms of the second and third generation
trilinear scalar couplings. In practice, because only third generation
Yukawa couplings are sizeable, the $A$-parameters of just the third
family are frequently relevant.

Although we have not shown this explicitly, the Yukawa couplings and the
$A$-parameters are, in general, (complex) matrices in generation
space. The resulting framework then requires $\ge 100$ parameters beyond
those of the SM\cite{dimop}, and hence is not very predictive. Since,
the phenomenology that we consider is generally insensitive to
inter-generation mixing of quarks and squarks, we assume that these
matrices are diagonal. Furthermore, since we do not discuss $CP$
violating effects, we take the superpotential and soft SUSY breaking
parameters to be real. Even so, a large number of additional parameters
remains. Most of these
occur in the SSB sector of the model, which simply reflects
our ignorance of the mechanism of supersymmetry breaking.  To
gain predictivity, despite the lack of a compelling model of SUSY
breaking, we must make additional simplifying assumptions about
symmetries of interactions at energy scales not directly accessible to
experiments, or postulate other physical principles that determine the
origin of the soft SUSY breaking terms.

The most popular model in which to embed the MSSM is the minimal
supergravity model (mSUGRA)~\cite{sugra}.  In this model, supersymmetry
is broken in a ``hidden sector'' which consists of fields which couple
to the fields of the visible sector (the MSSM fields) only
gravitationally.  Within the framework of supergravity grand
unification, the additional assumption that the vacuum expectation value
($vev$) of the gauge kinetic function does not break the unifying gauge
symmetry leads to a common mass $m_{1/2}$ for all gauginos.
%The assumption that  for the gauge kinetic function leads to
%a common mass $m_{1/2}$ for all gauginos.  
In addition, it is usually assumed that there exists a common mass $m_0$
for all scalars and a common trilinear term $A_0$ for all soft SUSY
breaking trilinear interactions.  Universal soft SUSY breaking scalar
masses are not, however, a consequence of the supergravity framework
\cite{soniweldon} but an additional assumption. 
%Flavor dependence of
%masses for sparticles with the same quantum numbers could lead flavor
%changing neutral currents (FCNC) at a level violating experimental
%constraints.  

The (universal) soft parameters are assumed to be renormalized at some
high scale $M_X \sim M_{GUT}-\mpl$. These are assumed to have values
comparable to the weak scale, $M_{weak}$, resulting in an elegant
solution to the fine-tuning problem associated with the Higgs
sector. Motivated by the apparently successful gauge coupling
unification in the MSSM, the scale $M_{GUT}\simeq 2\times 10^{16}$ GeV
is usually adopted for the scale choice $M_X$.  The resulting effective
theory, valid at energy scales $E<M_{GUT}$, is then just the MSSM with
soft SUSY breaking terms that unify at $M_{GUT}$.  The soft SUSY
breaking scalar and gaugino masses, the trilinear $A$ terms and in
addition a bilinear soft term $B$, the gauge and Yukawa couplings and
the supersymmetric $\mu$ term are all then evolved from $M_{GUT}$ to
some scale $M\simeq M_{weak}$ using renormalization group equations
(RGE). The large top quark Yukawa coupling causes the squared mass of
$H_u$ to be driven to negative values, which signals the radiative
breakdown of electroweak symmetry (REWSB); this then allows one to
determine the value of $\mu^2$ in terms of $M_Z^2$, possibly at the
expense of some fine-tuning. Finally, it is customary to trade the
parameter $B$ for $\tan\beta$, the ratio of Higgs field vacuum
expectation values. 
The resulting weak scale spectrum of superpartners and
their couplings can then be derived in terms of four continuous
parameters
plus one sign
\begin{equation}
m_0,\ m_{1/2},\ A_0,\ \tan\beta\ {\rm and}\ sign(\mu),
\label{msugra}
\end{equation}
in addition to the usual parameters of the standard model.  This
calculational procedure has been embedded into the event generator
ISAJET \cite{isajet} thereby allowing detailed predictions for the
collider events within this framework.
%expected from gravity-mediated (as well as gauge-mediated) SUSY breaking
%models.

The mSUGRA model, while highly 
predictive, 
%and dependent on only a rather small parameter set. 
%The simplicity of the mSUGRA model, however, is a consequence of 
rests upon a number of
simplifying assumptions that are invalid
in specific models of physics at energy scales $\sim M_{GUT}-\mpl$.
Thus, in the search for weak scale supersymmetry, the mSUGRA
model may give misleading guidance as to the possible event signatures
expected at high energy collider experiments. Indeed the literature
is replete with models with non-universal soft SUSY
breaking mass terms at the high scales.
In this paper, we survey a variety of these models 
(as well as others that lead to universality) and comment on 
possible phenomenological implications, especially for high energy collider
experiments. 
For the most part, we restrict our attention to models which reduce to the 
$R$-parity conserving MSSM at scales $Q<M_{GUT}$.

The event generator ISAJET (versions $>7.37$) has recently been 
upgraded~\cite{isajet} to accomodate supersymmetric models with non-universal
soft SUSY breaking masses at the $GUT$ scale. 
To generate such models, the user must input the usual mSUGRA
parameter set Eq. \ref{msugra}, but may in addition select one or 
several of the following options:
\begin{eqnarray*}
NUSUG1:& & M_1,\ M_2,\ M_3\\
NUSUG2:& & A_t,\ A_b,\ A_\tau \\
NUSUG3:& & m_{H_d},\ m_{H_u} \\
NUSUG4:& & m_{Q_1},\ m_{D_1},\ m_{U_1},\ m_{L_1},\ m_{E_1} \\
NUSUG5:& & m_{Q_3},\ m_{D_3},\ m_{U_3},\ m_{L_3},\ m_{E_3}.
\end{eqnarray*}
If one or more of the $NUSUGi$ ($i=1-5$) inputs are selected, then the
$GUT$ scale universal soft breaking masses are overwritten and a
weak-scale MSSM mass spectrum is generated.  ISAJET then computes the
corresponding branching fractions and sparticle cross sections, so that
specific theoretical predictions for $GUT$ scale SSB masses can be
mapped onto explicit predictions for the high energy collider events
expected to arise from these models.  In addition, the ISAJET keyword
$SSBCSC$ has been introduced in ISAJET versions $\ge 7.50$. Using
$SSBCSC$, the user may choose any scale between the weak scale and the
Planck scale at which to impose the above SSB boundary conditions. We
illustrate its use in Sec. XI C where it is necessary to introduce
boundary conditions at the string scale rather than at $M_{GUT}$.

To facilitate the examination of these models by our experimental
colleagues, we present here a survey of a number of well-motivated models
which usually lead to non-universality of SSB parameters.
Our survey is far from exhaustive, but is meant to present a flavor of the 
range of possibilities available for such models.
For each model, we
\begin{enumerate}
\item present a short description of the physics,
\item delineate the parameter space,
\item indicate how, within the model framework, collider events may be
generated using ISAJET, and
\item comment upon some of the general features of SUSY events expected
at collider experiments.
\end{enumerate}

The models selected include the following:
\begin{itemize}
\item $SU(5)$ grand unified models with universal soft SUSY breaking masses
at scales higher than $Q= M_{GUT}$,
\item $SU(5)$ models where supersymmetry breaking occurs via non-singlet
hidden sector superfields,
\item the MSSM plus an intermediate-scale right-handed neutrino which
leads to see-saw neutrino masses,
\item models with extra $D$-term contributions to scalar masses that are
generically present if the rank of the unifying gauge group exceeds 4,
\item minimal and general $SO(10)$ grand unified models with 
universal soft SUSY breaking masses at scales higher than $Q= M_{GUT}$,
\item grand unified models with group structure $G_{GUT}\times G_H$, 
where $G_H$ contains a hypercolor interaction used to solve the
doublet-triplet splitting problem,
\item effective supersymmetry models which lead to multi-TeV 
range scalar masses
for the first two generations, but sub-TeV masses for third generation
scalars and gauginos,
\item anomaly-mediated SUSY breaking models (AMSB), where the hidden sector
resides in different spacetime dimensions from the visible sector,
\item the minimal gaugino mediation model,
\item 4-dimensional string models with Calabi-Yao or orbifold
compactifications, and
\item models inspired by $M$-theory with SUSY breaking by one or several
moduli fields.
\end{itemize}
Space limitations preclude us from detailed discussions of these models.
Here, we sketch the physics behind each model, and provide the reader
with selected references where further details may be found. While much,
but by no means all, of the material presented may be found in the
literature, our hope is that the form in which we have presented it will
facilitate, or even spur, closer examination of alternatives to the
mSUGRA and gauge-mediated SUSY breaking models.

\section{$SU(5)$ grand unified model with the SSB universality scale  
higher than $M_{GUT}$}

As a working assumption, the scale at which all the SSB parameters are
generated, is usually taken
%, without good reason, 
to be $M_{GUT}$. If
this scale is substantially higher than this (but smaller than the
Planck scale), renormalization group (RG) evolution induces a
non-universality at the GUT scale.  The effect can be significant if
large representations are present.  Here, we assume that supersymmetric
$SU(5)$ grand unification is valid at mass scales $Q> M_{GUT}\simeq
2\times 10^{16}$ GeV, extending at most to the reduced Planck scale
$M_P\simeq 2.4\times 10^{18}$ GeV. Below $Q=M_{GUT}$, the $SU(5)$ model
breaks down to the MSSM with the usual $SU(3)_C\times SU(2)_L\times
U(1)_Y$ gauge symmetry. This framework is well described in, for
instance, the work of Polonsky and Pomarol\cite{pp}.

In the $SU(5)$ model, the $\hat{D}^c$ and $\hat{L}$ superfields are
elements of a $\bf{\bar{5}}$ superfield $\hat\phi$, while the $\hat{Q}$,
$\hat{U}^c$ and $\hat{E}^c$ superfields occur in the $\bf 10$
representation $\hat\psi$.  The Higgs sector is comprised of three
super-multiplets: $\hat{\Sigma} ({\bf 24})$ which is responsible for
breaking $SU(5)$, plus $\hat{\cal H}_1(\overline{\bf 5})$ and $\hat{\cal
H}_2({\bf 5})$ which contain the usual Higgs doublet superfields
$\hat{H_d}$ and $\hat{H_u}$ respectively, which occur in the MSSM. The
superpotential is given by,
\begin{eqnarray}
\hat{f}=\mu_\Sigma tr\hat{\Sigma}^2 &+&{1\over 6}\lambda'tr\hat{\Sigma}^3 +
\mu_H\hat{\cal H}_1\hat{\cal H}_2 +\lambda\hat{\cal H}_1\hat\Sigma
\hat{\cal H}_2 \\
&+&{1\over 4}f_t\epsilon_{ijklm}\hat{\psi}^{ij}\hat{\psi}^{kl}\hat{\cal H}_2^m 
+\sqrt{2}f_b\hat{\psi}^{ij}\hat{\phi}_i\hat{\cal H}_{1j} ,
\end{eqnarray}
where a sum over families is understood. $f_t$ and $f_b$ are the top and
bottom quark Yukawa couplings, $\lambda$ and $\lambda'$ are GUT Higgs sector
self couplings, and $\mu_\Sigma$ and $\mu_H$ are superpotential Higgs 
mass terms.
 
Supersymmetry breaking is parametrized by 
the soft supersymmetry breaking terms:
\begin{eqnarray}
{\cal L}_{soft}&=&-m_{{\cal H}_1}^2|{\cal H}_1|^2-
m_{{\cal H}_2}^2|{\cal H}_2|^2-
m_\Sigma^2tr\{ \Sigma^\dagger\Sigma \}-
m_5^2|\phi |^2 - m_{10}^2tr\{\psi^\dagger\psi \}
-{1\over 2}M_5\bar{\lambda}_{\alpha}
\lambda_{\alpha} \\
&+&\left[ B_\Sigma\mu_\Sigma tr\Sigma^2 +{1\over 6}A_{\lambda'}\lambda'
tr\Sigma^3 +B_H\mu_H{\cal H}_1{\cal H}_2 +A_\lambda\lambda{\cal H}_1\Sigma
{\cal H}_2 \right. \\
&+&\left. {1\over 4}A_tf_t\epsilon_{ijklm}\psi^{ij}\psi^{kl}{\cal H}_2^m
+\sqrt{2}A_bf_b\psi^{ij}\phi_i{\cal H}_{1j}+h.c.\right]
\end{eqnarray}

The various soft masses and gauge and Yukawa couplings evolve with 
energy according to the 15 renormalization group equations  
given in Appendix A of Ref. \cite{pp}. Here, we modify them to 
correspond with the sign conventions in ISAJET~\cite{isajet}:
\begin{eqnarray}
\frac{dm_{10}^2}{dt}&=&{1\over 8\pi^2}\left[ 3f_t^2 (m_{{\cal H}_2}^2+
2m_{10}^2+A_t^2)+2f_b^2(m_{{\cal H}_1}^2+m_{10}^2+m_5^2+A_b^2)-
{72\over 5}g_G^2M_5^2\right] , \\
\frac{dm_{5}^2}{dt}&=&{1\over 8\pi^2}\left[ 4f_b^2 
(m_{{\cal H}_1}^2+m_{10}^2+m_5^2+A_b^2)-{48\over 5}g_G^2M_5^2\right] ,\\
\frac{dm_{{\cal H}_1}^2}{dt}&=&{1\over 8\pi^2}\left[ 4f_b^2
(m_{{\cal H}_1}^2+m_{10}^2+m_5^2+A_b^2)+{24\over 5}\lambda^2
(m_{{\cal H}_1}^2+m_{{\cal H}_2}^2+m_\Sigma^2+A_\lambda^2)-
{48\over 5}g_G^2M_5^2\right] , \\
\frac{dm_{{\cal H}_2}^2}{dt}&=&{1\over 8\pi^2}\left[ 3f_t^2
(m_{{\cal H}_2}^2+2m_{10}^2+A_t^2)+{24\over 5}\lambda^2
(m_{{\cal H}_1}^2+m_{{\cal H}_2}^2+m_\Sigma^2+A_\lambda^2)-
{48\over 5}g_G^2M_5^2\right] , \\
\frac{dm_\Sigma^2}{dt}&=&{1\over 8\pi^2}\left[ {21\over 20}\lambda'^2
(3m_\Sigma^2+A_{\lambda'}^2)+\lambda^2(m_{{\cal H}_1}^2+m_{{\cal H}_2}^2+
m_\Sigma^2+A_\lambda^2)-
20g_G^2M_5^2\right] , \\
\frac{dA_t}{dt}&=&{1\over 8\pi^2}\left[ 9A_tf_t^2+4A_bf_b^2+
{24\over 5}A_\lambda \lambda^2+{96\over 5}g_G^2M_5\right] , \\
\frac{dA_b}{dt}&=&{1\over 8\pi^2}\left[ 10A_bf_b^2+3A_tf_t^2+
{24\over 5}A_\lambda \lambda^2+{84\over 5}g_G^2M_5\right] , \\
\frac{dA_\lambda}{dt}&=&{1\over 8\pi^2}\left[ {21\over 20}A_{\lambda'}
\lambda'^2+3A_tf_t^2+4A_bf_b^2+
{53\over 5}A_\lambda \lambda^2+{98\over 5}g_G^2M_5\right] , \\
\frac{dA_{\lambda'}}{dt}&=&{1\over 8\pi^2}\left[ {63\over 20}A_{\lambda'}
\lambda'^2+3A_\lambda \lambda^2+30g_G^2M_5\right] , \\
\frac{df_t}{dt}&=&{f_t\over 16\pi^2}\left[ 9f_t^2+4f_b^2+{24\over 5}\lambda^2
-{96\over 5}g_G^2\right] , \\
\frac{df_b}{dt}&=&{f_b\over 16\pi^2}\left[ 10f_b^2+3f_t^2+{24\over 5}\lambda^2
-{84\over 5}g_G^2\right] , \\
\frac{d\lambda}{dt}&=&{\lambda\over 16\pi^2}\left[ {21\over 20}\lambda'^2+
3f_t^2+4f_b^2+{53\over 5}\lambda^2
-{98\over 5}g_G^2\right] , \\
\frac{d\lambda'}{dt}&=&{\lambda'\over 16\pi^2}\left[ {63\over 20}\lambda'^2+
3\lambda^2-30g_G^2\right] , \\
\frac{d\alpha_G}{dt}&=& -3\alpha_G^2/{2\pi } , \\
\frac{dM_5}{dt}&=& -3\alpha_GM_5/{2\pi } ,
\end{eqnarray}
with $t=\log Q$.

To generate the weak scale MSSM mass spectrum, one begins with
the input parameters
\begin{equation}
\alpha_{GUT},\ f_t,\ f_b,\ \lambda ,\ \lambda'
\label{y_su5}
\end{equation}
stipulated at $Q=M_{GUT}$, where $f_b=f_\tau$ is obtained from the
corresponding mSUGRA model.  The first three of these
can be extracted, for instance, from ISASUGRA, versions $\ge 7.44$.  The
couplings $\lambda (M_{GUT})$ and $\lambda' (M_{GUT})$ are additional
inputs, where $\lambda (M_{GUT})\agt 0.7$~\cite{pdk} to make the triplet
Higgsinos heavy enough to satisfy experimental bounds on the proton
lifetime. The gauge and Yukawa couplings can be evolved via the RGEs to
determine their values at $Q=M_P$.  Assuming universality at $M_P$ (this
maximizes the effects of non-universality at the GUT scale), one imposes
\begin{eqnarray}
m_{10}&=&m_5=m_{{\cal H}_1}=m_{{\cal H}_2}=m_\Sigma \equiv m_0 \\
A_t&=&A_b=A_\lambda =A_\lambda'\equiv A_0,
\label{MpBoundary}
\end{eqnarray}
and evolves all the soft masses from $M_P$ to $M_{GUT}$.
The MSSM soft breaking masses at $M_{GUT}$ are specified via
\begin{eqnarray}
&m_Q^2=m_U^2=m_E^2\equiv m_{10}^2\,,&
\nonumber\\
&m_D^2=m_L^2\equiv m_5^2\,,&
\label{MGUTboundary}\\
&m_{H_1}^2=m_{{\cal H}_1}^2\,,\quad m_{H_2}^2=m_{{\cal H}_2}^2\,,&
\nonumber
\end{eqnarray}
which can serve as input to ISAJET~\cite{isajet} via the $NUSUGi$ keywords.
Since there is no splitting amongst the gaugino masses, the
gaugino masses may be taken to be $M_1=M_2=M_3\equiv m_{1/2}$ where
$m_{1/2}$ is stipulated most conveniently at the $GUT$ scale.

To obtain correct Yukawa unification, it is crucial to start with the
correct weak scale Yukawa couplings.  To calculate the values of the
Yukawa couplings at scale $Q=M_Z$, one begins with the pole masses
$m_b=4.9$ GeV and $m_\tau =1.784$ GeV.  One may calculate the
corresponding running masses in the $\overline{MS}$ scheme, and evolve
$m_b$ and $m_\tau$ up to $M_Z$ using 2-loop SM RGEs.  At $Q=M_Z$, the
SUSY loop corrections to $m_b$ and $m_\tau$ must be included; ISAJET
versions $>7.44$ uses the approximate formulae of Pierce {\it et
al.}\cite{pierce}.  A similar procedure is used to calculate the top
quark Yukawa coupling at scale $Q=m_t$. SUSY particle mass spectra
consistent with constraints from collider searches and with unified $b$
and $\tau$ Yukawa couplings (to 5\%) are then obtained (assuming
universality of scalar masses at the scale $M_P$), but only for $\mu <0$
and $30 \alt \tan\beta \alt 50$, where the allowed range is weakly
sensitive to $\alpha_s$.

To illustrate the extent of non-universality due to $SU(5)$ running of
SSB masses between $M_P$ and $M_{GUT}$, we explicitly examine a typical
case. The corresponding input parameters as well as the values of SSB
parameters at $M_{GUT}$ are listed in Table \ref{tsu5_1}.  The GUT scale
input parameters extracted from ISAJET for $\tan\beta =35$ are
$f_t=0.534$ and $f_b=f_\tau =0.271$ for the top, bottom and tau Yukawa
couplings.  We also adopt $\lambda=1.0$ and $\lambda'=0.1$ for the
$SU(5)$ Higgs couplings and $g_{GUT}=0.717$ for the unified $SU(5)$
gauge coupling.  At the Planck scale, we then take $m_0=150$ GeV and
$A_0=0$ GeV, parameters that are analogous to $m_0$ and $A_0$ at the GUT
scale in the mSUGRA model.  We take $m_{1/2}(M_{GUT})=200$ GeV for the
universal gaugino masses.

The evolution of SUSY mass parameters in the minimal $SU(5)$ model
between $M_P$ and $M_{GUT}$ is shown in Fig.~\ref{mpmg1}, assuming
universality at $M_P$.  We see that the rather high value of $\lambda$
induces a large splitting $m_5^2\simeq m_{10}^2>m_{{\cal H}_1}^2,\
m_{{\cal H}_2}^2$.  Likewise, the large value of $f_t$ is responsible
for the splitting $m_{{\cal H}_1}^2 > m_{{\cal H}_2}^2$ at $M_{GUT}$.
The large $t$ and $b$ Yukawa couplings are also responsible for the
split between third generation and the first two generation values of
$m_{10}$ and $m_5$.  It is interesting to notice that reasonable values
of the free parameters can give $\sim 100$\% deviations from
universality at $M_{GUT}$. In the cases that we checked, it was typically
the Higgs scalars that are split by the large amount from the other
scalars, primarily because $\lambda$ is large: for acceptable solutions,
the corresponding non-universality between matter scalar masses was
typically $\sim 10-20$\%. In Table \ref{tsu5_2}, we list the corresponding
values of selected weak scale sparticle masses for both the $SU(5)$ case
and mSUGRA. The shift in scalar masses in this case can be up to $\sim
20\%$, with the biggest shift occuring in the $\tell_R$ and $\ttau_1$
masses.

\section{$SU(5)$ models with non-universal gaugino masses.}

Since supergravity is not a renormalizable theory, in general we may
expect a non-trivial gauge kinetic function, and hence the possiblity of
non-vanishing gaugino masses if SUSY is broken. Expanding the gauge
kinetic function as
$f_{ab} = \delta_{ab} + \hat{\Phi}_{ab}/\mpl + \ldots$, where
the fields
$\hat{\Phi}_{ab}$ transform as left handed chiral superfields
under supersymmetry transformations, and as
the symmetric product of two adjoints under gauge symmetries, we parametrize
the lowest order contribution to gaugino masses by,
\begin{equation}
{\cal L}\supset \int d^2\theta {\hat{f}}^a{\hat{f}}^b
{\hat{\Phi}_{ab}\over M_{\rm Planck}} + h.c.
\supset  {\langle F_{\Phi} \rangle_{ab}\over M_{\rm Planck}}
\lambda^a\lambda^b\, +\ldots ,
\end{equation}
where the $\lambda^a$ are the gaugino fields,
and $F_{\Phi}$ is the auxillary field component of $\hat{\Phi}$ that
acquires a SUSY breaking $vev$.

If the fields $F_{\Phi}$ which break supersymmetry are gauge singlets,
universal gaugino masses result.  However, in principle, the chiral
superfield which communicates supersymmetry breaking to the gaugino
fields can lie in any representation in the symmetric product of two
adjoints, and so can lead to gaugino mass terms\footnote{The results of
this section are not new, but in the interest of completeness we thought
it fit to include a review of these models in this section.} that
(spontaneously) break the underlying gauge symmetry. We require, of
course, that SM gauge symmetry is preserved.
Non-universal gaugino masses have been previously considered by other
authors\cite{Hill,ellis,drees,anderson}.  

In the context of $SU(5)$
grand unification, $F_{\Phi}$ belongs to an $SU(5)$ irreducible
representation which appears in the symmetric product of two adjoints:
\begin{equation}
({\bf 24}{\bf \times}
 {\bf 24})_{\rm symmetric}={\bf 1}\oplus {\bf 24} \oplus {\bf 75}
 \oplus {\bf 200}\,,
\label{irrreps}
\end{equation}
where only $\bf 1$ yields universal masses.  
The relations amongst the
various GUT scale gaugino masses have been worked out {\it e.g.} in
Ref. \cite{anderson}.  The relative $GUT$ scale $SU(3)$, $SU(2)$ and
$U(1)$ gaugino masses $M_3$, $M_2$ and $M_1$ are listed in
Table~\ref{masses} along with the approximate masses after RGE evolution
to $Q\sim M_Z$.  Here, motivated by the measured values of the
gauge couplings at LEP, we assume that the $vev$ of the SUSY-preserving
scalar component
of ${\hat{\Phi}}$ is neglible. 
Each of the three non-singlet models is as predictive as the
canonical singlet case, and all are compatible with the unification of
gauge couplings.  These scenarios represent the predictive subset
of the more general (and less predictive) case of an arbitrary superposition
of these
representations.
The model parameters may be chosen to be, 
\begin{equation}
m_0,\ M_3^0,\ A_0,\ \tan\beta\ {\rm and}\ sign(\mu ),
\end{equation}
where $M_i^0$ is the $SU(i)$ gaugino mass at scale $Q=M_{GUT}$.  $M_2^0$
and $M_1^0$ can then be calculated in terms of $M_3^0$ according to
Table \ref{masses}.
Sample spectra for each case are exhibited in Table \ref{tnusug}.

The phenomenology of these models has recently been examined in
Ref. \cite{nusug}, and the SUSY reach presented for Fermilab
Tevatron upgrade options for a variety of discovery
channels. The results were found to be model-dependent.
In particular, in the {\bf 24} model, a large splitting
between weak scale values of $m_{\tz_2},\ m_{\tw_1}$ and $m_{\tz_1}$
gave rise to large rates for events with isolated leptons, so that SUSY
discovery should be easier in this case than in the mSUGRA model.  A
special feature of this model is the sizeable cross section for
$(Z\to\ell\bar{\ell})+jets+\eslt$ events. Indeed, for certain ranges of
model parameters, SUSY discovery seemed to be
possible only via this channel.
In contrast, for the {\bf 75} and {\bf 200} models, $m_{\tz_2}$,
$m_{\tw_1}$ and $m_{\tz_1}$ were all nearly degenerate, so that leptons
arising from --ino decays were very soft and difficult to detect.
Consequently, there was hardly any reach for SUSY in these models at the
Tevatron via leptonic channels, and the best reach occurred typically in
the $\eslt +jets$ channels.

\section{The MSSM with a right handed neutrino}

Experimental evidence~\cite{oscill} strongly indicates the existence of
neutrino oscillations, and almost certainly neutrino mass. 
The favoured interpretation is $\nu_{\mu}-\nu_{\tau}$ oscillations, with
$\Delta m^2 \sim 10^{-2}$~eV$^2$ and near-maximal mixing. An attractive
method for introducing neutrino mass into the MSSM is via the see-saw
mechanism\cite{seesaw}. In this case, one can introduce an additional
chiral superfield\footnote{Our purpose here is to
illustrate the effect of introducing singlet neutrino superfields on
the SSB parameters and the SUSY spectrum. An explanation of the
atmospheric neutrino data would, of course, require us to introduce more
than one such superfield and also interactions that violate lepton
flavour conservation, but as long as these have only small
Yukawa couplings, their effect on the spectrum should be negligible.}
($\hat{N}^c$) which transforms as a gauge singlet
(whose fermionic component is the left-handed anti-neutrino and scalar
component is $\tnu_R^{\dagger}$). A Majorana
mass term for the right-handed neutrino is allowed and, because $\nu_R$ is
a SM singlet, its mass may be large: $M_N\sim 10^{10}-10^{16}$ GeV. When
electroweak symmetry is broken, a Dirac neutrino mass $m_D \sim
m_{\ell}$ is also induced via the usual Higgs mechanism. The resulting
neutrino mass matrix must be diagonalized, and one obtains a light
physical neutrino mass $m_\nu\simeq m_D^2/M_N$ plus a dominantly
singlet neutrino of mass $M\simeq M_N$.

The superpotential for the MSSM with a singlet neutrino 
superfield $\hat{N}^c$ (for
just a single generation), is given by
\begin{equation}
\hat{f}=\hat{f}_{MSSM}+f_{\nu}\epsilon_{ij}\hat{L}^i\hat{H}_u^j\hat{N}^c
+{1\over 2}M_N\hat{N}^c\hat{N}^c
\label{WmssmN}
\end{equation}
while the soft SUSY breaking terms now include
\begin{equation}
{\cal L}={{\cal L}_{MSSM}}-m_{\tnu_R}^2 |\tnu_R |^2 +\left[
A_{\nu}f_{\nu}\epsilon_{ij}\tilde{L}^i\tilde{H}_u^j\tilde{\nu}_R^\dagger
+ \frac{1}{2} B_{\nu} M_N \tnu_R^2 + h.c. \right].
\label{LmssmN}
\end{equation}
The parameters $A_{\nu}$, $B_{\nu}$
and $m_{\tnu_R}$ are assumed to be comparable to the weak scale.

 Many of the relevant RGEs have been presented in Ref.\cite{rhn}. Here
we present the complete set needed for determining the sparticle
spectrum at the weak scale.
The one-loop RGEs for the gauge couplings and gaugino masses are 
unchanged from the MSSM case, since the $\hat{N}^c$ superfield is a gauge
singlet.
The Yukawa coupling RGEs are
\begin{eqnarray}
{{df_t}\over{dt}}&=&{{f_t}\over{16\pi^2}}\bigg[6f_t^2+f_b^2+f_{\nu}^2
-{{16}\over 3}g_3^2-3g_2^2-{{13}\over{15}}g_1^2\bigg]
\label{htRN}\\
{{df_b}\over{dt}}&=&{{f_b}\over{16\pi^2}}\bigg[f_t^2+6f_b^2+f_{\tau}^2
-{{16}\over 3}g_3^2-3g_2^2-{7\over{15}}g_1^2\bigg]
\label{hbRN}\\
{{df_{\tau}}\over{dt}}&=&{{f_{\tau}}\over{16\pi^2}}\bigg[
3f_b^2+4f_{\tau}^2+f_{\nu}^2-3g_2^2-{9\over 5}g_1^2\bigg]
\label{htauRN}\\
{{df_{\nu}}\over{dt}}&=&{{f_{\nu}}\over{16\pi^2}}\bigg[3f_t^2+f_{\tau}^2
+4f_{\nu}^2-3g_2^2-{3\over 5}g_1^2\bigg] .
\label{hnRN}
\end{eqnarray}
%
%For the RGEs of the scalar masses, we assume that the 
%hypercharge trace ${\rm{tr}}(Y'm^2)$ is zero. 
The RGEs for $m_Q^2$, $m_U^2$, $m_D^2$, $m_E^2$ and $m_{H_d}^2$ are 
all unchanged from the MSSM. 
However, for $m_L^2$, $m_{\tnu_R}^2$ and $m_{H_u}^2$, we have
\begin{eqnarray}
{{dm_L^2}\over{dt}}&=&{2\over{16\pi^2}}\Big[-{3\over 5}g_1^2M_1^2
-3g_2^2M_2^2+f_{\tau}^2X_{\tau}
+f_{\nu}^2X_n\Big]
\label{mLRN}\\
{{dm_{\tnu_R}^2}\over{dt}}&=&{4\over{16\pi^2}}\Big[f_{\nu}^2X_n\Big]
\label{mNRN}\\
{{dm_{H_u}^2}\over{dt}}&=&{2\over{16\pi^2}}\Big[-{3\over 5}g_1^2M_1^2
-3g_2^2M_2^2
+3f_t^2X_t+f_{\nu}^2X_n\Big]
\label{mH2RN}
\end{eqnarray}
where we have defined $X_n=m_L^2+m_{\tnu_R}^2+m_{H_u}^2+A_\nu^2$
and $X_t$ and $X_\tau$ are given in Ref. \cite{bbo}.
Finally, the RGEs for the $A_i$ parameters are given by
\begin{eqnarray}
{{dA_t}\over{dt}}&=&{2\over{16\pi^2}}\Big[\Sigma c_ig_i^2M_i +6f_t^2A_t+
f_b^2A_b+f_{\nu}^2 A_{\nu}\Big] \\
{{dA_b}\over{dt}}&=&{2\over{16\pi^2}}\Big[\Sigma c_i'g_i^2M_i +6f_b^2A_b+
f_t^2A_t+f_\tau^2A_\tau\Big] \\
{{dA_\tau}\over{dt}}&=&{2\over{16\pi^2}}\Big[\Sigma c_i''g_i^2M_i +3f_b^2A_b+
4f_\tau^2 A_\tau+f_{\nu}^2A_{\nu}\Big] \\
{{dA_{\nu}}\over{dt}}&=&{2\over{16\pi^2}}\Big[\Sigma c_i'''g_i^2M_i+ 
3f_t^2A_t+4f_{\nu}^2A_{\nu}+f_\tau^2 A_\tau\Big] ,
\end{eqnarray}
where the $c_i$, $c_i'$ and $c_i''$ are given in Ref.~\cite{bbo}, and
$c_i'''=\{{3\over 5},3,0\}$. These RGEs apply for scales $Q>M_N$, while
the MSSM RGEs are used below $Q=M_N$.  Below the scale $M_N$ the
effective theory does not contain the right handed neutrino or
sneutrino, so that the running of the corresponding parameters is frozen
at their values at this scale.  The RGE for the parameter $B_{\nu}$ is
irrelevant for our analysis.

This model has been explicitly included in ISAJET version $\geq 7.48$, via the
keyword $SUGRHN$, which allows, in addition to $mSUGRA$ and/or $NUSUGi$
inputs, the following:
\begin{equation}
m_{\nu_\tau},\ M_N,\ m_{\tnu_{\tau R}},\ A_{\nu} ,
\end{equation}
where all masses are entered in GeV units. Then the neutrino Yukawa 
coupling is calculated, and the MSSM+RHN RGEs are used at scales
$Q>M_N$, while MSSM RGEs are used below $Q=M_N$.

A sample spectrum of masses is shown in Table \ref{trhn}, assuming
$m_{\nu_\tau}=10^{-9}$~GeV, $M_N=10^{13}$~GeV, $m_{\tnu_{\tau R}}=200$~GeV
and $A_{\nu}=0$. The main effect is that the additional Yukawa coupling
drives the third generation slepton masses to somewhat lower values
than the massless neutrino case. 

An upper limit on the parameter $\tan\beta$ occurs in mSUGRA for
$\mu <0$ due to
a breakdown in the REWSB mechanism, where the $H_u$ mass is not
driven sufficiently negative by RG running. For the MSSM+RHN model,
the additional Yukawa coupling $f_\nu$ aids somewhat in driving $m_{H_u}^2$
negative. 
It is natural to ask how much the additional Yukawa coupling $f_\nu$
would help to increase the allowed range
of $\tan\beta$ while still satisfying the REWSB constraint. As an
example, we checked that for the case $m_0=m_{1/2}=200$~GeV, $A_0=0$,
and $\mu<0$, for which $\tan\beta \le 45$ in the mSUGRA framework, the
inclusion of a right-handed neutrino with $m_N=10^{13}$, $(10^{10})$
$((10^7))$~GeV, only increases this range to 45.3 (45.7) ((46)), assuming
$f_{\nu}=f_t$ at the $GUT$ scale.

\section{Unifying gauge groups with rank $\ge 5$: D-terms}

In general, if the MSSM is embedded in a $GUT$ gauge group with rank
$\ge 5$, and the $GUT$ gauge group is spontaneously broken to a gauge
group of lower rank, there are additional $D$-term contributions to
scalar masses.  The important thing is that these
contributions~\cite{dreesDterm} affect TeV scale physics even if the scale
at which the $GUT$ symmetry is broken is very large: since symmetry
breaking is arranged to occur in a nearly $D$-flat direction, these
$D$-term contributions to scalar masses are still of order the weak
scale, even though the extra particles have masses $\sim M_{GUT}$.  The
$D$-terms must be added to the various SUSY scalar mass squared parameters at
the high scale at which the breaking occurs, so that these effectively
lead to non-universal boundary conditions for scalar masses.

Kolda and Martin\cite{kolda} have analysed these contributions for 
gauge groups which are subgroups of $E_6$, which encompasses a 
wide range of well-motivated $GUT$ group choices. $E_6$ contains in addition
to the SM $SU(3)_C\times SU(2)_L\times U(1)_Y$ gauge symmetry two additional
$U(1)$ symmetries labelled as $U(1)_X$ and $U(1)_S$. 
The $D$-term contributions to scalar masses can then be parametrized as,
\begin{eqnarray}
\Delta m_Q^2&=&{1\over 6}D_Y-{1\over 3}D_X-{1\over 3}D_S, \nonumber \\
\Delta m_D^2&=&{1\over 3}D_Y+D_X-{2\over 3}D_S, \nonumber \\
\Delta m_U^2&=&-{2\over 3}D_Y-{1\over 3}D_X-{1\over 3}D_S, \nonumber \\
\label{dterms}
\Delta m_L^2&=&-{1\over 2}D_Y+D_X-{2\over 3}D_S \\
\Delta m_E^2&=&D_Y-{1\over 3}D_X-{1\over 3}D_S, \nonumber \\
\Delta m_{H_d}^2&=&-{1\over 2}D_Y-{2\over 3}D_X+D_S, \nonumber \\
\Delta m_{H_u}^2&=&{1\over 2}D_Y+{2\over 3}D_X+{2\over 3}D_S \nonumber,
\end{eqnarray}
where $D_Y$ is the usual $D$-term associated with weak hypercharge
breaking.  In light of our ignorance of the mechanism of gauge symmetry
breaking, the contributions $D_X$ and $D_S$ can be treated as additional
dimensionful parameters, that can range over positive as well as negative
values.

\subsection{Minimal $SO(10)$ model with gauge symmetry breaking at $Q=M_{GUT}$}

A simple special case of the above arises if the $GUT$ gauge group
$SO(10)$ is assumed to directly break to the SM gauge group at
$Q=M_{GUT}$ so the theory below this scale is the MSSM, possibly
together with a right-handed neutrino and sneutrino. In this case,
the three generations of matter superfields plus an additional SM gauge
singlet right handed neutrino superfield for each generation are each
elements of the 16 dimensional spinor representation of $SO(10)$, and
so are taken to have a common mass $m_{16}$ above $M_{GUT}$.  The Higgs
superfields of the MSSM belong to a single 10 dimensional fundamental
representation of $SO(10)$, and acquire a mass $m_{10}$.  At
$Q=M_{GUT}$, the gauge symmmetry breaking induces
\begin{equation}
D_X\ne 0;\ \ \ D_Y=D_S=0
\end{equation}
so that at this scale the scalar masses are broken according to (\ref{dterms}).
The MSSM masses at $M_{GUT}$ may then be written as
\begin{eqnarray}
m_Q^2=m_E^2=m_U^2=m_{16}^2+M_D^2 \nonumber \\
m_D^2=m_L^2=m_{16}^2-3M_D^2 \\
m_{H_{u,d}}^2=m_{10}^2\mp 2M_D^2, \nonumber
\end{eqnarray}
where we have reparametrized $D_X=-3M_D^2$. If the right-handed neutrino
mass is substantially below the GUT scale, the soft breaking sneutrino
mass would evolve as in Eq.~(\ref{mNRN}); at the $GUT$ scale it
would then be given by, 
\begin{equation}
m_{\tnu_R}^2 = m_{16}^2+5M_D^2.
\end{equation}

In {\it minimal} $SO(10)$, the superpotential above $M_{GUT}$ has the
form,
\begin{eqnarray}
\hat{f} = f\hat{\psi}\hat{\psi}\hat{\phi} +\cdots
\label{spsoten}
\end{eqnarray}
with just a single Yukawa coupling per generation, where $\hat{\psi}$
and $\hat\phi$ are the $\bf 16$ dimensional spinor and $\bf 10$
dimensional Higgs superfields, respectively. 
We neglect possible inter-generational mixing and also assume that the
right-handed neutrino has a mass $\sim M_{GUT}$.
The dots represent terms including for instance higher dimensional
Higgs representations and interactions responsible for the breaking
of $SO(10)$. We assume here for simplicity that these couplings are 
suppressed relative to the usual Yukawa couplings.

In minimal $SO(10)$, all the Yukawa couplings are unified above $M_{GUT}$, 
which
forces one into a region of very large $\tan\beta\sim 50$ which is
actually excluded assuming universality of scalars 
if the constraint of radiative electroweak symmetry breaking is included. 
It has been suggested\cite{sotendterms}, and recently shown\cite{bdft}, 
that $D$-term contributions have the correct form to allow for
Yukawa unified solutions to the SUSY particle mass spectrum 
consistent with radiative electroweak symmetry breaking. 

The parameter space of this model
can be taken as
\begin{equation}
m_{16},\ m_{10},\ M_D^2,\ m_{1/2},\ A_0,\ sign(\mu ),
\end{equation}
where $M_D^2$ can be either positive or negative.
Yukawa
coupling unification forces $tan\beta$ to be in the range 45-52 --
for many purposes its exact value is irrelevant.

The parameter space of minimal $SO(10)$ SUSY GUT models was 
explored in Ref. \cite{bdft}. It was found that,
requiring Yukawa unification good to 5\%, {\it no}
solutions could be found for values
of $\mu >0$, while many solutions could be obtained for $\mu<0$, 
but only for {\it positive} values of $M_D^2$. The $D$-term forces
$m_{H_u}<m_{H_d}$ at $Q=M_{GUT}$: this is necessary to drive 
$m_{H_u}^2$ negative before $m_{H_d}^2$, as is required for REWSB with
$\tan\beta > 1$.
Implications of this model for the dark matter relic density,
$b\to s \gamma$ decay rate, and collider searches, are presented in 
Ref. \cite{so10_2}

A sample spectrum from the mSUGRA model and a corresponding case in
Yukawa-unified $SO(10)$ are shown in Table \ref{tso10_1}.  The $D$-term
splitting that ameliorates REWSB also leaves a distinct imprint on the
masses of the matter scalars: the left- sleptons and right- down-type
squarks have smaller GUT scale squared masses than their
counterparts. This can be reflected in the weak scale spectrum where
left- sleptons can be lighter than right- sleptons, and the right bottom
squark can be by far the lightest of all the squarks -- perhaps, even
within the kinematic reach of the Main Injector upgrade of the Tevatron,
though its detection may be complicated. Note also the smaller absolute
value of the $\mu$ parameter in the $SO(10)$ case: this results in
lighter charginos and neutralinos with substantial, or even dominant,
higgsino components and a smaller $\tz_2-\tz_1$ mass difference. Finally,
we remark that for the case shown, the lighter $\ttau$ is dominantly
$\ttau_L$. 

It is well known\cite{bbct} that SUSY models with $\mu< 0$ and large
$\tan\beta$ yield a large rate for the decay $b\to s\gamma$. 
Indeed\cite{so10_2}, this class of models is already severely constrained by
experimental results on radiative $b$-decays. 
However, additional non-universality between generations is possible
in this framework, which could alter the gluino loop contributions, and
hence the final branching fraction for $b\to s\gamma$ decay.

\section{Mass splittings in $SO(10)$ above $Q=M_{GUT}$}

\subsection{Minimal $SO(10)$}

As discussed above, the minimal $SO(10)$ model contains three generations
of matter superfields each in a {\bf 16} dimensional representation, and
a {\it single} Higgs superfield in the {\bf 10} dimensional
representation. The superpotential is as given in Eq.~(\ref{spsoten})
with $f$ the common Yukawa coupling for the third generation.
Other terms will also be present, including Yukawa couplings for the
first two generations, as well as more complicated Higgs representations
necessary for $SO(10)$ breaking. We will assume the Yukawa couplings
involving these fields are all small, so the dominant contribution to
RGE running comes from just the superpotential (\ref{spsoten}). We also
assume associated $SO(10)$ soft SUSY breaking parameters: $m_{16}$,
$m_{10}$, $m_{1/2}$ and $A$.  Then the RGEs in the minimal $SO(10)$
model are calculable.  For the gauge coupling we have,
\begin{equation}
{{dg}\over{dt}}={g^3\over{16\pi^2}}(S-24),
\label{gRGE}
\end{equation}
where $S$ is the sum of Dynkin indices of the various 
chiral superfields in the model.
With the above minimal field content, $S=7$. However, additional fields
associated for instance with $SO(10)$ breaking ought to be present, and 
will increase the value of $S$.
The Yukawa coupling RGE is,
\begin{equation}
{{df}\over{dt}}={1\over{16\pi^2}}f
\Big(14f^2-{\textstyle{63\over 2}}
g^2\Big) .
\label{hRGE}
\end{equation}
For the gaugino mass we have the following RGE:
\begin{equation}
{{dm_{1/2}}\over{dt}}={1\over{16\pi^2}}2(S-24)g^2m_{1/2}
\label{MhalfRGE}
\end{equation}
For the scalar masses we have:
\begin{eqnarray}
{{dm_{16}^2}\over{dt}}&=&{1\over{16\pi^2}}\bigg[
10f^2\Big(2m_{16}^2+m_{10}^2+A^2\Big)-45g^2m_{1/2}^2\bigg]
\label{m16RGE}\\
{{dm_{10}^2}\over{dt}}&=&{1\over{16\pi^2}}\bigg[
8f^2\Big(2m_{16}^2+m_{10}^2+A^2\Big)-36g^2m_{1/2}^2\bigg] .
\label{mHRGE}
\end{eqnarray}
Finally, the RGE for the trilinear mass parameter is
\begin{equation}
{{dA}\over{dt}}={1\over{16\pi^2}}\Big(28f^2 A+63g^2m_{1/2}\Big) .
\label{AtRGE}
\end{equation}

As an illustration, we adopt the minimal $SO(10)$ case 5 spectra from
Ref. \cite{bdft} for which Yukawa couplings unify at $M_{GUT}$.  The
model parameters and mass spectrum is listed in the ``$M_{GUT}$
Unification'' column of Table \ref{msoten}.  We begin by using 
$f(M_{GUT})=0.553$ and $g_{GUT}=0.706$ (as given by the minimal $SO(10)$
model).  We then evolve (using $S=7$) from $M_{GUT}$ to $M_P$ to find
the corresponding Planck scale gauge and Yukawa couplings. At $M_P$, we
assume universality of the three generations with $m_{16}=629.8$ GeV,
while $m_{10}=836.2$ GeV.  At $M_{GUT}$, we take $m_{1/2}=348.8$ GeV and
$A=-186.5$ GeV, with a $D$-term $M_D=135.6$ GeV.  A Yukawa unified
solution is obtained for $\tan\beta =52.1$ and the corresponding
spectrum is shown in the last column titled ``$M_P$ Unification''.

In Fig. \ref{so10_min}, we show by the solid lines the effect of running
of SSB parameters between $M_P$ and $M_{GUT}$ for the minimal $SO(10)$
model, for parameters as in Table \ref{msoten}.
The dashed lines show the corresponding situation for $S=15$,
{\it i.e. with one additional adjoint included}; in this case, the
running of the gauge coupling between $M_{GUT}$ and $M_P$ (see
Eq.~\ref{gRGE}) is somewhat slower. We see that the splitting $\delta
m_{16}^2$ between the GUT scale mass parameters of the first (or second)
and third generations is reduced, albeit by a small
amount.~\footnote{Since the right hand side of Eq.~(\ref{hRGE}) is more
negative when $S=15$ as compared to the $S=7$ case, the corresponding
$f$ runs to smaller values in the former case. If we now consider
the evolution of $\delta m_{16}^2$, for which the term depending on $g$
drops out, we see that this difference runs the most for $S=7$ for which
$f$ is largest.
In this sense, the difference shown by the solid
lines may be regarded as a bound.}

The effect of $SO(10)$ running is that the first two generations of
matter scalars run to higher masses, while the Higgs masses and third
generation masses decrease somewhat. 
The corresponding weak
scale sparticle masses are listed in Table \ref{msoten}, without and
with the effect of Planck to $GUT$ scale running. The main effect is a
$\sim 27$\% change in the mass difference between the (lightest) charged
sleptons of the first and third generations.

\subsection{General $SO(10)$}

More generally, we may take the two MSSM Higgs doublets to live in different
fundamental representations of $SO(10)$: $\hat{H}_u\in \hat{H}_2$ and
$\hat{H}_d\in \hat{H}_1$. Then the superpotential can be written as
\begin{equation}
\hat{f}=f_t\hat{\psi}\hat{\psi}\hat{H}_2 +
f_b\hat{\psi}\hat{\psi}\hat{H}_1 ,
\label{spgso10}
\end{equation}
so that there exist two Yukawa couplings above the GUT scale, 
and just $f_b=f_\tau$ unification, which can 
occur for a much wider range of $\tan\beta$ values\cite{pierce}, is required.
In addition to the usual scalar masses, as in Ref. \cite{bagger}, 
we include an off-diagonal mass term $m_{H_{12}}^2({H}_1^\dagger
{H}_2 +{H}_2^\dagger{H}_1)$. As in minimal $SO(10)$, there 
should also be at least higher dimensional Higgs representations
present responsible for $SO(10)$ breaking, but again, we ignore these.

We give here the RGEs for the general $SO(10)$ model, 
thereby completing the results of Refs. \cite{bhs,bagger}. 
For the gauge coupling constant we have:
\begin{equation}
{{dg}\over{dt}}={{g^3}\over{16\pi^2}}(S-24) ,
\label{gRGEso10}
\end{equation}
where $S$ again is the sum of the Dynkin indices of the $SO(10)$
fields. 
For just two 10 dimensional Higgs multiplets and
3 generations of matter, $S=8$. 
For gaugino masses, we again have
\begin{equation}
{{dm_{1/2}}\over{dt}}={1\over{16\pi^2}}2(S-24)g^2m_{1/2} .
%\label{MhalfRGE}
\end{equation}
The Yukawa coupling RGEs are:
\begin{eqnarray}
{{df_t}\over{dt}}&=&{{f_t}\over{16\pi^2}}\left(
14f_t^2+14f_b^2-{{63}\over 2}g^2\right)
\label{htRGESO10}\\
{{df_b}\over{dt}}&=&{{f_b}\over{16\pi^2}}\left(
14f_t^2+14f_b^2-{{63}\over 2}g^2\right) .
\label{hbRGESO10}
\end{eqnarray}
The RGEs for the scalar masses are given by
\begin{eqnarray}
{{dm_{16}^2}\over{dt}}&=&{10\over{16\pi^2}}
\left[ f_t^2(2m_{16}^2+m_{H_2}^2)+f_b^2(2m_{16}^2+m_{H_1}^2)+
2f_tf_b m_{H_{12}}^2\right.\nonumber\\
&&\left.+(A_t^2f_t^2+A_b^2f_b^2)-{9\over 2} g^2m_{1/2}^2\right]
\label{m16RGESO10}\\
{{dm_{H_1}^2}\over{dt}}&=&{8\over{16\pi^2}}\Big[f_b^2(2m_{16}^2+m_{H_1}^2)
+f_tf_b m_{H_{12}}^2+A_b^2f_b^2-{9\over 2}g^2m_{1/2}^2\Big]
\label{mH1RGESO10}\\
{{dm_{H_2}^2}\over{dt}}&=&{8\over{16\pi^2}}\Big[f_t^2(2m_{16}^2+m_{H_2}^2)
+f_tf_b m_{H_{12}}^2+A_t^2f_t^2-{9\over 2}g^2m_{1/2}^2\Big]
\label{mH2RGESO10}\\
{{dm_{H_{12}}^2}\over{dt}}&=&{4\over{16\pi^2}}\Big[
f_tf_b(4m_{16}^2+m_{H_1}^2+m_{H_2}^2+2A_tA_b)+
(f_t^2+f_b^2)m_{H_{12}}^2\Big] .
\label{mH12RGESO10}
\end{eqnarray}
Finally, the RGEs for the $A$ parameters are
\begin{eqnarray}
{dA_t\over dt}&=&{1\over{16\pi^2}}(28f_t^2A_t+20f_b^2A_b
+63g^2 m_{1/2})\\
{dA_b\over dt}&=&{1\over{16\pi^2}}(28f_b^2A_b+20f_t^2A_t
+63g^2 m_{1/2}).
\end{eqnarray}
We show in Fig. \ref{so10_gen} the running of SSB parameters in the
general $SO(10)$ model using $GUT$ scale values of $g=0.717$,
$f_t=0.534$ and $f_b= 0.271$, as in Fig. \ref{mpmg1}. Except for
$m_{H_{12}}^2$ which is fixed to be zero at $Q=M_P$, the SSB parameters
are also as in this figure.  The main effect is again a significant
splitting between first or second and third generation scalar masses at
the $GUT$ scale. Some splitting between $m_{H_u}$ and $m_{H_d}$ also occurs,
with $m_{H_u}^2 < m_{H_d}^2$ as desired.  The corresponding weak scale
sparticle masses are shown in Table \ref{gsoten}. The $GUT$ scale SSB
term splitting results in somewhat heavier scalars than in the $mSUGRA$
case.  For this example, because most of the weak scale squark mass
comes from the RG evolution, the effect is more pronounced for sleptons
than squarks. In particular, this increase is just a few percent for
squarks, but as much as 22\% for sleptons. 
%Nonetheless, it is unlikely
%that the determination of sparticle masses alone can serve to
%distinguish the two frameworks.

\section{Supersymmetric missing partner models with hypercolor}

In this variety of models, the gauge group is of the type $G_{GUT}\times
G_H$, where the first group is $SU(5)$ or $SO(10)$ and the second is
related to a `hypercolor' interaction \cite{hypercolor,HIY}. While the
weak $SU(2)$ is completely contained in the first factor, colour $SU(3)$
is not embedded in either of the factors.  Although the gauge group is
not simple, an approximate unification of the gauge coupling constants
of the group $SU(3)_C\times SU(2)\times U(1)$ is achieved if the
couplings of $G_H$ are large enough.  These models provide a solution to
the doublet-triplet splitting problem by the missing partner mechanism.
Since the MSSM gauginos do not belong to a single multiplet of a simple
gauge group, their masses do not obey the usual unification condition
\cite{ACM}, resulting in non-universality of gaugino masses. However, if usual
squarks and sleptons and the MSSM Higgs fields
are singlets of $G_H$, universality of scalar masses is still possible,
as for instance, in the $SU(5)_{GUT}\times SU(3)_H \times U(1)_H$ model of
Ref.~\cite{HIY}, where the hypercharge $U(1)$ is a combination of
$U(1)_H$ and a $U(1)$ subgroup in the first factor.

In this case, the following relations among gauge couplings hold
at the unification scale \cite{ACM}
\begin{equation}
{1\over{g_1^2}}={1\over{g_{GUT}^2}}+{1\over{15g_{H1}^2}}\,,\qquad
{1\over{g_2^2}}={1\over{g_{GUT}^2}}\,,\qquad
{1\over{g_3^2}}={1\over{g_{GUT}^2}}+{1\over{g_{H3}^2}}\,,
\label{GaugatGUT}
\end{equation}
where $\sqrt{3/5}g_1$, $g_2$, and $g_3$ are the gauge couplings of the
$U(1)_Y$, $SU(2)_L$, and $SU(3)_C$ SM groups, and $g_{GUT}$, $g_{H3}$,
and $g_{H1}$ are the $SU(5)_{GUT}$, $SU(3)_H$, and $U(1)_H$ unified
groups respectively.  Clearly from Eq.~(\ref{GaugatGUT}) we see that the
unification of the gauge coupling constants from low energy data is
achieved if $g_{H1}^2\gg g_{GUT}^2$ and $g_{H3}^2\gg g_{GUT}^2$. In
addition, considering that the prediction for $\alpha_s$ at the weak
scale in SUSY GUT models (without threshold corrections) is higher than
the world averaged experimental value, it was argued that the correction
introduced by hypercolor moves the prediction for $\alpha_s$ in the
correct direction. It was found that \cite{ACM}:
\begin{equation}
\alpha_s(m_Z)\approx 0.130 -{{0.014}\over{\alpha_{H3}}}-
{{0.010}\over{15\alpha_{H1}}}
\label{alphastrong}
\end{equation}
where $\alpha_i=g_i^2/4\pi$ and threshold corrections have been neglected.
In order for $\alpha_s$ not to shift too much, we must have
$\alpha_{H3}\agt 0.6$ and $\alpha_{H1}\agt 0.03$, though for
$\alpha_{H1}$ as small as 0.03, $g_2^2-g_1^2 = 0.18g_1^2 g_2^2$.

Above the GUT scale there are three gauginos associated to the groups
$SU(5)_{GUT}$, $SU(3)_H$, and $U(1)_H$ whose masses we denote $m_{1/2}$,
$M_{H3}$ and $M_{H1}$ respectively. Below the GUT scale we have the MSSM
and the three MSSM gauginos are a linear combination of the former ones.
The masses of the bino, wino, and gluino are then given by
\footnote{A somewhat different model\cite{hypercolor,HIY} based on the group
$SO(10)_{GUT}\times SO(6)_H$ also has non-universal MSSM gaugino masses.
However, since the hypercolor group is simple, there is one relation
between them\cite{kurosawa}},
\begin{eqnarray}
M_1&=&g_1^2\left({{m_{1/2}}\over{g_{GUT}^2}}+{{M_{H1}}\over{15g_{H1}^2}}
\right)\,,\nonumber\\
M_2&=&m_{1/2}\,, \label{GaugMass}\\
M_3&=&g_3^2\left({{m_{1/2}}\over{g_{GUT}^2}}+{{M_{H3}}\over{g_{H3}^2}}
\right)\,.\nonumber
\end{eqnarray}
The thing to note is that $M_{H1,3}/\alpha(H_{1,3})$ are renormalization
group invariants (at one loop) so that $M_{H1,3}/g_{H1,3}^2$ need not be
small even when $g_{H1,3}^2$ is large.  The relative magnitude of the
three masses $m_{1/2}$, $M_{H3}$ and $M_{H1}$ is unknown because it
depends on the SUSY breaking mechanism.  One might naively suppose that
they are of the same order of magnitude; in this case, gaugino masses
could be significantly different at the GUT scale, though the magnitude
of the non-universality would be limited because, as noted above, the
couplings $g_{H1}$ and $g_{H3}$ have to be considerably larger than
$g_{GUT}$. There is no reason, however, why $M_{H1}$ and $M_{H3}$ cannot
be much larger than $m_{1/2}$. Indeed in scenarios with dilaton
dominated SUSY breaking, we have\cite{dilaton,ibanez1}
\begin{equation}
{m_{1/2} \over g_{GUT}^2} = {M_{H1} \over g_{H1}^2} = {M_{H3} \over g_{H3}^2},
\label{eq:dilaton}
\end{equation}
so that gaugino mass splittings of O(100\%) are expected.

In Fig.~\ref{hyperc}{\it a} we plot non--universal gaugino masses of the MSSM
as a function of a common hypercolor gaugino mass $M_{H1}=M_{H3}\equiv M_H$.
We take $m_{1/2}=200$ GeV, $g_{GUT}=0.716$, and two different choices for
the hypercolor gauge couplings: $\alpha_{H1}=0.1$ and $\alpha_{H3}=0.7$
in solid lines, and $\alpha_{H1}=0.5$ and $\alpha_{H3}=0.8$ in dashed
lines. As indicated in Eq.~(\ref{GaugMass}) the wino mass $M_2$ is always
equal to $m_{1/2}=200$ GeV. The other two gaugino masses are larger
(smaller) than $M_2$ if the hypercolor gaugino mass is larger (smaller)
than $m_{1/2}$. The gluino mass deviates more from $M_2$ compared to the
bino mass because of the factor 15 in Eq.~(\ref{GaugMass}) and our
choice of values for other parameters.
%because
%the hypercolor gauge coupling $g_{H3}$ has been taken larger than $g_{H1}$.
The larger the hypercolor gauge couplings, the smaller the deviations from
universality. In addition, if the common hypercolor mass is equal to
$m_{1/2}$ there is no deviation from universality no matter the value of the
hypercolor gauge couplings. In Fig.~\ref{hyperc}{\it b}, we show the
same gaugino masses but assuming instead that $M_{H1}/g_{H1}^2 =
M_{H_3}/g_{H3}^2$. The three gauge couplings are chosen exactly as in
frame {\it a}) so that there is a large hierarchy between the masses of
the gauginos of the three groups. The cross denotes the
dilaton-dominated scenario for which point Eq.~(\ref{eq:dilaton})
is satisfied. Indeed we see that very large non-universality of gaugino
masses may be possible.

In Fig. \ref{hyper1}, we show several weak scale sparticle masses versus
the same parameter $M_H$ as in Fig. \ref{hyperc} for parameter values
corresponding to the solid curves in this figure.  The two frames
illustrate the results for the same choices of the gaugino masses as in
Fig.~\ref{hyperc}.  In frame {\it a}) we see that the non-colored
sparticle masses hardly vary at all versus $M_H$, while the gluino and
squark masses can vary by up to 12\%. This is presumably because the
coloured sparticle masses run considerably more than those of uncoloured
sparticles coupled with the fact that $M_3$ varies more with $M_H$ than
$M_1$ does, and $M_2$ does not change at all.  The variation is, of
course, much more dramatic in frame {\it b}). For very large values of
$M_{H1}$, the coloured sparticles as well as the heavier chargino and
neutralinos become very heavy, and may be in conflict with fine-tuning
considerations. We also mention that although $M_1$ starts out larger
than $M_2$ at the GUT scale (but not by a huge amount), ${M_1 \over
M_2}$ is driven to a value close to $1 \over 2$ at the weak scale for
acceptable values of $M_{H1}$: it would be interesting to examine
whether precise measurements of masses and mixing angles could lead to
observable deviations from expectations in mSUGRA or gauge-mediated SUSY
breaking frameworks. In the same vein, we also mention that $m(\te_R)$
also increases slowly from 132~GeV in mSUGRA to 134~GeV for the dilaton
dominated scenario to 151~GeV for the extreme case with $M_{H1}=3$~TeV,
while $m(\te_L)$ is roughly constant. This is because the RG evolution
of $m(\te_R)^2$ is due to hypercharge gauge interactions, and $M_1$
starts out bigger than $M_2$ (which is independent of $M_{H1}$).

\section{Models with effective supersymmetry}

The SM exhibits accidental global symmetries which inhibit
flavor--changing neutral currents (FCNC), lepton flavor violation (LFV),
electric dipole moments (EDM) of electron and neutron, and proton decay,
as opposed to the MSSM where degeneracy or alignment in the mass
matrices has to be invoked.  On the other hand, supersymmetry stabilizes
the scalar masses under radiative corrections, contrary to the SM where
it is hard to understand the hierarchy between the Higgs mass and the
Planck scale. The models presented in this section \cite{DKS,CKN} aim to
combine the good features of both the SM and the MSSM. There are two
mass scales: gauginos, higgsinos, and third generation squarks are
sufficiently light ($\alt 1$ TeV) to naturally stabilize the Higgs mass
and the electroweak scale, while the first two generations of squarks
and sleptons (whose Yukawa couplings to Higgs are very small) are
sufficiently heavy ($\widetilde M\sim5$ to 20 TeV) to suppress FCNC,
LFV, {\it etc.}. This class of models, called Effective
Supersymmetry, 
does not invoke degeneracy or alignment in the mass matrices.

In one of the realizations of Effective Supersymmetry\cite{CKN}, the
first two generations of squarks and sleptons, together with the
down--type Higgs, are composite, with constituents that carry a
``superglue'' charge, and have a mass $\sim\widetilde M$. Gauge
superfields, third generation superfields and the up--type
Higgs superfield are taken to be fundamental and neutral under
superglue, with perturbative couplings to the constituents, so that
their mass is
suppressed relative to the mass of the composites.  In this way,
the spectrum is characterized as follows.
\begin{itemize}
\item Gaugino masses are light and can be non-universal with masses given
by $M_i=n_i(\alpha_i/4\pi)\widetilde M$, where $n_i$ are numerical factors
that can be as large as ${\cal O}(10)$. 
\item Left and right squark and slepton masses for the first two
generations are of the order of $\widetilde M$.
\item Left and right squark and slepton masses for the third generation 
are of the order of $(\lambda_3/4\pi)\widetilde M$; for $\lambda_3 \sim
1$, this is an  
order of magnitude smaller than $\widetilde M$.
\item The down--Higgs mass satisfy $m_{H_d}\sim\widetilde M$. The 
up--Higgs mass on the other hand, is given by
$m_{H_u}\sim (\lambda_H/4\pi)\widetilde M$, where $\lambda_H$ is its
perturbative coupling to the constituents. Therefore,
there is only one Higgs in the low energy theory and 
$\tan\beta\sim4\pi/\lambda_H$ is large.
\item The ``$\mu$--term'' and the ``$B\mu$--term'' respectively satisfy 
$\mu\sim(\lambda_H/4\pi)\widetilde M$ and 
$B\mu\sim(\lambda_H/4\pi)\widetilde M^2$.
\end{itemize}
%
%In addition, to provide the correct electroweak scale we need
To obtain $m_{H_u} \sim$~100~GeV, we require
$\lambda_H/4\pi\sim10^{-2}$, while 
$\lambda_3/4\pi\sim10^{-1}$ ensures $m_{\tilde t} \alt 1$~TeV.

If the hierarchy of scalar masses is already present at the unification
scale, then it has been shown that unless the stop mass squared at the
unification scale is taken to be well above (1~TeV)$^2$, two-loop
contributions to scalar renormalization group equations drive the top
squark mass squared negative well before the weak scale, resulting in a
breakdown of color symmetry\cite{murayama}. Thus, this simple class of
models seems to be ruled out by fine-tuning considerations. 
To account for this class of constraints, we have implemented
the full set of two-loop MSSM RGEs in ISAJET versions $\ge 7.50$.

Very
recently, Hisano {\it et al.} \cite{HKN} have identified scenarios in
which first and second generation scalars can be much heavier than
gauginos and scalars of the third generation, and for which the scalar
masses are renormalization group invariant (so that the
constraints of Ref.\cite{murayama} are not relevant) as long as gaugino
masses are neglected in the RGEs.  These constraints
are also inapplicable in models in which the assumption of the scalar
hierarchy is made for mass parameters at a scale $\sim 10-50$~TeV, since
then there
are no large logs that drive $m_{\tt}^2$ to negative values. In this
case, however, model-dependent {\it finite} contributions to $\delta
m_{\tt}^2$ are no longer negligible, and need to be examined to discuss
the viability of any particular model \cite{AG}.

%However, interest in models with a hierarchy of scalar masses at the
%unification scale has recently been revived
Yet another possibility has been considered in Ref.\cite{feng,bagger,bagger2}.
%by Feng {\it et al.}\cite{feng} and Bagger {\it et al.}\cite{bagger}.
These authors begin with {\it all} scalar masses initially at the
multi-TeV scale at or above $M_{GUT}$, and show that for certain choices
of $M_{GUT}-\mpl$ scale boundary conditions on the scalar masses and $A$
parameters-- keeping gaugino masses at the weak scale-- the third generation
sfermion and Higgs masses are driven to weak scale values, while scalars
of the first two generations remain heavy. Such a scenario is
particularly attractive in the context of minimal $SO(10)$. In this
case, with Yukawa coupling unification plus a singlet $\hat{N}^c$,
particularly simple boundary conditions \cite{bagger2},
\begin{equation}
%m_U^2=m_Q^2=m_D^2=m_L^2=m_E^2=m_N^2={1\over 2}m_{H_u}^2={1\over 2} m_{H_d}^2
%={1 \over 4}A^2
4m_{16}^2 = 2m_{10}^2 = A^2
\label{boundary}
\end{equation} 
lead to sub-TeV scale third generation scalar masses, while first and
second generation scalar masses can be as high as 20~TeV. If instead the
boundary value of $A$ is taken to be at the weak scale, the hierarchy
generated\cite{bagger} is somewhat smaller.  Examples of sparticle mass
spectra were not generated in Ref. \cite{bagger}, where it was noted
that this scenario shares the problem of obtaining correct radiative
breaking of electroweak symmetry common to most high $\tan\beta$
scenarios: in examples shown in Ref. \cite{bagger} and
Ref. \cite{bagger2}, the two Higgs SSB masses stay positive at all
scales in their evolution to the weak scale, with $m_{H_u}>m_{H_d}$,
contrary to what is needed for REWSB. 

In a recent analysis \cite{imh} it has been shown that if the boundary
conditions in Eq.~(\ref{boundary}) are augmented by $SO(10)$ $D$-terms,
it is possible to obtain the desired inverted mass hierarchy amongst the
squarks together with
radiative electroweak symmetry breaking. This
then yields a calculable model based on the gauge group $SO(10)$ with
(approximate) unification of Yukawa couplings. 
The analysis in Ref.\cite{imh} took the
right-handed neutrino mass to be fixed near $\sim 10^{13}$~GeV,
and obtained ``crunch'' factor values $S$ up to $\sim 5-7$
for full $SO(10)$ $D$-terms, and factors of $S$ up to
9 if splittings were applied only to the soft SUSY breaking
Higgs masses.
The crunch factor $S$ is defined as
\begin{eqnarray*}
S   =
 {{3(m_{u_L}^2+m_{d_L}^2+m_{u_R}^2+m_{d_R}^2)+m_{\te_L}^2+m_{\te_R}^2+
m_{\tnu_e}^2}
 \over 
{3(m_{\tst_1}^2+m_{\tb_1}^2+m_{\tst_2}^2+
m_{\tb_2}^2)+m_{\ttau_1}^2+m_{\ttau_2}^2+
m_{\tnu_{\tau}}^2}}.
\end{eqnarray*}
These values are considerably below those quoted in Ref. \cite{bagger2},
where a more idealized case was considered.

Effective supersymmetry is not as mature a framework as mSUGRA or the
gauge-mediated SUSY breaking. Except for the inverted hierarchy model of
the previous paragraph, all the models discussed in this Section suffer
from incompleteness which preclude computations at as thorough a level.
The scenario in Ref.\cite{CKN} involves new unknown strong dynamics at
the 10~TeV scale. Models where the splitting between third generation
scalars and those of the other generations has a dynamical origin
\cite{feng,bagger,bagger2} suffer from the fact that this dynamics does not
break electroweak symmetry: the mass spectrum thus does
not appear to be calculable unless deviations such as non-universality
are imposed.
These considerations notwithstanding, collider events for generic
effective SUSY models can be generated with ISAJET~\cite{isajet} by using the 
weak-scale $MSSMi$
keywords, with independent weak scale SSB masses as inputs. One may
enter multi-TeV scale first and second generation scalar masses, while
using sub-TeV scale gaugino masses, third generation scalar masses and
$\mu$ parameters. In the scenario of Ref.\cite{CKN}, $A$-terms are
${\cal O}(100)$~GeV or smaller, while $m_A$ is very large. 

Sparticle mass spectra from the radiatively generated
inverted mass hierarchy solution due to
Bagger {\it et al.} are not possible without modifications that allow
REWSB to occur. 
Two possibilities are the non-universalities due to $SO(10)$ $D$-terms,
or ad-hoc Higgs sector splittings. These may be implemented
in ISAJET using the  NUSUG inputs along with the right-handed
neutrino solution. In ISAJET, if a zero physical neutrino mass is
entered, then the Yukawa couplings $f_t$ and $f_\nu$ automatically
unify.
It remains to be seen whether the resulting inverted mass hierarchy is
truly sufficient to solve problems due to FCNCs, LFVs and
the EDM of the electron and neutron.

\section{Anomaly-mediated SUSY breaking}

In most models, soft SUSY breaking parameters of the low energy
effective theory are thought to receive contributions from gravitational
or gauge interactions which are considered to be messengers of SUSY
breaking in a hidden sector. It has recently been recognized
\cite{RanSun,grlm} that there is an additional contribution, that
originates in the super-Weyl anomaly, which is always present when SUSY
is broken. In models without SM gauge singlet superfields that can
acquire a Planck scale $vev$, the usual supergravity contribution to
gaugino masses is suppressed by an additional factor $\frac{M_{SUSY}}
{M_P}$ relative to $m_{\frac{3}{2}} = M_{SUSY}^2/M_P$, and the
anomaly-mediated contribution can dominate. These contributions are
determined in terms of the SUSY breaking scale by the corresponding
$\beta$ functions.

\begin{equation}
M_i={\beta_g \over g}m_{\frac{3}{2}},
\label{GaugMassSeq}
\end{equation}
where $\beta_i$ is the one--loop beta function, defined by
$\beta_{g_i}\equiv dg_i/d\ln\mu=-b_ig_i^3+...$. The gaugino masses are
not universal, but given by the ratios of the respective $\beta$-functions.

In general, however, K\"ahler potential couplings between the observable
sector and the hidden sector (Goldstino) field, which are generically
not forbidden by a symmetry, result in large gravity contributions
($\sim m_{\frac{3}{2}}$) to scalar masses which would completely
dominate the corresponding anomaly-mediated contributions. These gravity
contributions can be strongly suppressed if the SUSY breaking and
visible sectors reside on different branes, and are ``sufficiently
separated'' in a higher dimensional space: in this case, the suppression
is the result of geometry and not a symmetry, though then one has to
wonder about the dynamics that results in such a geometry.
The anomaly-mediated contribution is given by,
\begin{equation}
m_{\tilde q}^2=-\quarter\left\{{{d\gamma}\over{dg}}\beta_g+
{{d\gamma}\over{df}}\beta_f\right\}m_{\frac{3}{2}}^2
%\half\left\{cbg^4-dy^2(ey^2+fg^2)\right\}m_{\frac{3}{2}}^2,
\label{ScalMassSeqGen}
\end{equation}
where $\beta_g$ and $\beta_f$ are the $\beta$ functions for gauge and
Yukawa interactions, respectively, and 
$\gamma=\partial\ln Z/\partial\ln\mu$, with $Z$
the wave function renormalization constant.  Notice that this is
comparable to the corresponding contribution to the gaugino
masses. Furthermore, since Yukawa interactions are negligible for the
first two generations, the anomaly-mediated contributions to scalar
masses of the first two generations are essentially
equal. Unfortunately, however \cite{RanSun}, the anomaly contribution
turns out to be negative for sleptons, necessitating additional sources
for the squared masses of scalars. There are several proposals in the
literature, but phenomenologically it suffices to add a universal
contribution $m_0^2$ (which, of course, preserves the degeneracy between
the first two generations of scalars) to Eq.~(\ref{ScalMassSeqGen}), and
regard $m_0$ as an additonal parameter\cite{wells}. 

Finally, in the sign convention of ISAJET~\footnote{This is
opposite to that used in Ref.~\cite{wells}.}, the anomaly-mediated
contribution to the trilinear SUSY breaking scalar coupling is given by,
\begin{equation}
A_f=+{\beta_f \over f} m_{\frac{3}{2}}.
\label{ASeq}
\end{equation}
It is assumed that the {\it ad hoc} introduction of $m_0^2$ in
Eq.~(\ref{ScalMassSeqGen}) does not affect the other relations.

\subsection{The Minimal Anomaly-Mediated SUSY Breaking Model (AMSB)}

In this framework, it is assumed that the  anomaly-mediated SUSY
breaking contributions to the soft-SUSY breaking contributions dominate,
and further, that the introduction of the parameter $m_0^2$ is sufficent
to circumvent the problem of negative squared masses for sleptons. 
The parameter space of the model consists of
\begin{equation}
m_0,\ m_{3/2},\ \tan\beta\ {\rm and}\ sign(\mu ) .
\end{equation}
In this case, gaugino masses are given by
\begin{eqnarray}
M_1&=& {33\over 5}{g_1^2\over 16\pi^2}m_{3/2} ,\\
M_2&=& {g_2^2\over 16\pi^2}m_{3/2} ,\ {\rm and} \\
M_3&=& -3{g_3^2\over 16\pi^2}m_{3/2} .
\end{eqnarray}
Third generation scalar masses are given by
\begin{eqnarray}
m_{U}^2 &=& \left(-{88\over 25}g_1^4+8g_3^4+2f_t\hat{\beta}_{f_t}\right)
{m_{3/2}^2\over (16\pi^2)^2}+m_0^2,\\
m_{D}^2 &=& \left(-{22\over 25}g_1^4+8g_3^4+2f_b\hat{\beta}_{f_b}\right)
{m_{3/2}^2\over (16\pi^2)^2}+m_0^2,\\
m_{Q}^2 &=& \left(-{11\over 50}g_1^4-{3\over 2}g_2^4+
8g_3^4+f_t\hat{\beta}_{f_t}+f_b\hat{\beta}_{f_b}\right)
{m_{3/2}^2\over (16\pi^2)^2}+m_0^2,\\
m_{L}^2 &=& \left(-{99\over 50}g_1^4-{3\over 2}g_2^4+
f_\tau\hat{\beta}_{f_\tau}\right) {m_{3/2}^2\over (16\pi^2)^2}+m_0^2,\\
m_{E}^2 &=& \left(-{198\over 25}g_1^4+
2f_\tau\hat{\beta}_{f_\tau}\right) {m_{3/2}^2\over (16\pi^2)^2}+m_0^2,\\
m_{H_u}^2 &=& \left(-{99\over 50}g_1^4-{3\over 2}g_2^4+
3f_t\hat{\beta}_{f_t}\right) {m_{3/2}^2\over (16\pi^2)^2}+m_0^2,\\
m_{H_d}^2 &=& \left(-{99\over 50}g_1^4-{3\over 2}g_2^4+
3f_b\hat{\beta}_{f_b}+f_\tau\hat{\beta}_{f_\tau}\right) 
{m_{3/2}^2\over (16\pi^2)^2}+m_0^2 .
\end{eqnarray}
The $A$-parameters are given by
\begin{eqnarray}
A_t&=&{\hat{\beta}_{f_t}\over f_t}{m_{3/2}\over 16\pi^2},\\
A_b&=&{\hat{\beta}_{f_b}\over f_b}{m_{3/2}\over 16\pi^2},\ {\rm and} \\
A_\tau &=&{\hat{\beta}_{f_\tau}\over f_\tau}{m_{3/2}\over 16\pi^2}.
\end{eqnarray}
In the above, we have
\begin{eqnarray}
\hat{\beta}_{f_t} &=& 16\pi^2\beta_t=f_t\left( -{13\over 15}g_1^2-3g_2^2
-{16\over 3}g_3^2+6f_t^2+f_b^2\right),\\
\hat{\beta}_{f_b} &=& 16\pi^2\beta_b=f_b\left( -{7\over 15}g_1^2-3g_2^2
-{16\over 3}g_3^2+f_t^2+6f_b^2+f_\tau^2\right),\\
\hat{\beta}_{f_\tau} &=& 16\pi^2\beta_\tau=f_\tau
\left( -{9\over 5}g_1^2-3g_2^2+3f_b^2+4f_\tau^2\right).
\end{eqnarray}
The first two generations of squark and slepton masses are given by the 
corresponding formulae above with the Yukawa couplings set to zero.
This model has been implemented in ISAJET versions $\ge 7.45$,
using the $AMSB$ keyword, which allows input of the above
parameter space set. In $ISAJET$, it is easiest to implement the above 
masses at scale $Q=M_{GUT}$, and proceed with evolution to the weak scale.
Then the $B$ and $\mu^2$ parameters are calculated in accord with 
the constraint from radiative electroweak symmetry breaking.

The most notable feature of this framework is the hierarchy of gaugino
masses. The gluino is (as in mSUGRA) much heavier than the electroweak
gauginos, but the novel feature is that ${M_1 \over M_2} \sim 3.2$, so
that the wino is by far the lightest supersymmetric particle (LSP). The
wino LSP scenario has several implications for phenomenology, the most
important of which is the near degeneracy of the chargino and the
(wino-like) neutralino LSP.  One loop corrections\cite{pp,su,wells,cdm},
which make the dominant contribution to the chargino-neutralino mass gap,
have been included~\cite{isajet} in ISAJET v7.46 (in the gaugino limit).  The
phenomenology can be sensitive to this mass difference\cite{su,wells}.

In Table \ref{tamsb}, we show spectra generated from the minimal 
AMSB model for
two values of $m_0$, with other parameters being the same. Note that the
parameter $m_{3/2}$ should be selected typically above 25,000 GeV to
avoid constraints from LEP experiments.
>From the spectra shown, we immediately 
see several well-known aspects of the $AMSB$ spectrum. Most notably, 
we see that the $\tw_1$ and $\tz_1$ are nearly degenerate in
mass, so that in addition to the usual leptonic decay modes $\tw_1 \to
\tz_1\ell\nu$, the only other allowed (and in these cases dominant)
decay of the chargino is $\tw_1^{\pm} \to \tz_1\pi^{\pm}$.
%only decay mode allowed for $\tw_1$ is $\tw_1\to e\nu_e \tz_1$.
The chargino has a very small width, corresponding to a lifetime 
$\sim 1.5 \times 10^{-9}$~s, so that it would be expected to travel a
significant fraction of a meter before decaying \cite{wells}.
Secondly, the $\tell_L$ and $\tell_R$ are nearly mass degenerate.  This
degeneracy (which seems fortuitous) is much tighter than expected in the
mSUGRA framework and certainly in the gauge-mediated SUSY breaking
framework.  Their mass scale is largely determined by the parameter
$m_0$, and it is possible that for small enough $m_0$ slepton signals
may be detectable at the next generation of $e^+e^-$ colliders or even
at the LHC. Another interesting feature (which may serve to distinguish
the cases shown from mSUGRA) is that the $\ttau_L-\ttau_R$ mixing is
near maximal. The prospects for measuring this have been discussed in
Ref.\cite{nojiri}.

In the minimal AMSB framework, $m_{\tw_1}-m_{\tz_1}$ is typically bigger than
160~MeV, so that $\tw_1\to \tz_1 \pi$ is always allowed and the
chargino typically decays within the detector~\cite{wells}.
%decay length of a few cm. or less, so that the $\tw_1$ will decay within
%a collider detector into a very soft charged pion. 
The chargino would then manifest itself only as missing energy, unless
the decay length is a few tens of cm, so that the chargino track can be
established in the detector. The track would then seem to
disappear~\cite{su} since the presence of the soft pion would be very
difficult to detect. Some parameter regions with $m_{\tw_1} - m_{\tz_1}
< m_{\pi^{\pm}}$ may be possible; in this case, the chargino would
mainly decay via $\tw_1 \to \tz_1 e \nu$ and its decay length (depending
on the mass difference) would be typically larger than several metres. It
would then show up via a search for long-lived charged exotics.

There have been a number of alternative suggestions to cure the
negative slepton mass squared problem\cite{amsbprime}. 
Generally, these require
the introduction of additional fields at energy scales higher
than the weak scale. The mass spectrum in these scenarios differs from that
of the minimal AMSB model sketched above, and characteristic features
such as $m_{\tw_1}\simeq m_{\tz_1}$ and $m_{\tell_L}\simeq m_{\tell_R}$
need not occur. These models are not hard wired into ISAJET, but 
can be generated using the $NUSUG_i$ inputs at a scale dictated by $SSBCSC$;
in this case, the user must perform the calculation of the SSB masses of
MSSM particles.

\section{Minimal Gaugino Mediation}

Very recently, Schmaltz and Skiba\cite{mgm} have proposed a model based
on extra dimensions with branes, which is claimed to provide novel
solutions to the SUSY flavour and $CP$ problems. Within their framework,
chiral supermultiplets of the observable sector reside on one brane
whereas the SUSY breaking sector is confined to a different
brane~\cite{RanSun}. Gravity and gauge superfields propagate in the
bulk, and hence, directly couple to fields on both the branes. As a
result of their direct coupling to the SUSY breaking sector, gauginos
acquire a mass. The scalar components of the chiral supermultiplets,
however, can acquire a SUSY breaking mass only via their interactions
with gauginos (or gravity) which feel the effects of SUSY breaking: as a
result, these masses are suppressed relative to gaugino masses, and may
be neglected in the first approximation. The same is true for the $A$-
and $B$-parameters.

In the specific realization\cite{mgm}, to preserve the success of the
unification of gauge couplings, it is assumed that there is grand
unification (both $SU(5)$ and $SO(10)$ are discussed), and further, that
the compactification scale $M_c$ below which there are no Kaluza-Klein
excitations, is larger than $M_{GUT}$. Furthermore, since light bulk
fields have flavor-blind interactions by construction, it is argued
that the scale $M_c \alt M_{Planck}/10$ in order to sufficiently
suppress flavour violating scalar couplings (due to heavy bulk fields)
that would be generically present. Based on the discussion in the
previous paragraph, they take the boundary conditions for the soft SUSY
breaking parameters of the MSSM to be, $m_0=A_0=B_0=0$ at the scale
$M_c$, and argue that the spectrum is completely specified by the
parameter set,
\begin{equation}
\mu , m_{1/2}, M_c
\label{mgm1}
\end{equation}
where it is the grand unification assumption that leads to a universal
gaugino mass above $Q=M_{GUT}$. They refer to this as the Minimal
Gaugino Mediation (MGM) model. The parameters $m_{1/2}$ and $\mu$
should be comparable, and are chosen to be $\sim M_{Weak}$.
The REWSB constraints fix $\mu^2$, while the requirement $B_0=0$ fixes
$\tan\beta$. In Ref.~\cite{mgm} it is shown that if $M_c \leq M_{Planck}/10$ 
$\tan\beta$ lies between $\sim 12$ and $\sim 18$
(12-25) for the $SU(5)$ ($SO(10)$) model with ${\bf {5}}+ {\bf
\bar{5}}$ (${\bf {16}}+ {\bf \bar{16}}$) Higgs supermultiplets in addition
to the usual adjoint Higgs multiplet. The LSP may be the stau,
the lightest neutralino or the gravitino. However, the latter has a weak
scale mass, and as in the mSUGRA framework, is irrelevant for collider
phenomenology. 

Our purpose here is to outline how to generate sample spectra in this
framework using ISAJET~\cite{isajet}, and examine some issues that have
not been discussed in Ref. \cite{mgm}. For definiteness, we will choose
the $GUT$ group to be $SU(5)$. This model is then a special case of our
discussion in Sec.~II, except that the SSB parameters now ``unify'' at
the scale $M_c$ rather than $M_P$ (where they take on the special
values). Our first observation is that the allowed range of $\tan\beta$
seems incompatible\cite{pierce} with $\tan\beta \geq 30$ required for
the unification of the $\tau$ and $b$ Yukawa couplings\footnote{Another
possibility is the inclusion of a bilinear $R$-parity violating term in
the tau sector. In this case, $b$-$\tau$ Yukawa unification can be
achieved at smaller values of $\tan\beta$\cite{dfrv}.}. For this reason,
and also because the prediction for $\tan\beta$ could depend on how the
$\mu$ problem might be solved, we will ignore the $B_0 = 0$ condition
and treat $\tan\beta$ as a phenomenological
parameter.\footnote{Moreover, if Higgs fields are also allowed to
propogate in the bulk \cite{higgsbulk}, we would expect $B_0 \sim
m_{1/2} \sim m_{H_u} \sim m_{H_d}$.} For our analysis, we modify the
model parameters~\footnote{There are other coupling constants involving
GUT scale physics, but we will see that these do not significantly
change the spectrum.} to,
\begin{equation}
m_{1/2}, M_c, \tan\beta , sign(\mu ).
\label{mgm2}
\end{equation}
As before, the user will have to obtain the values of the SSB parameters
at $Q=M_{GUT}$ using the RG equations of Sec.~II, and input these into
ISAJET for generating mass spectra and/or collider events as desired. As
shown in Table \ref{mgmtable}, we fix $m_{1/2}$ at the GUT scale and
$\tan\beta$ at the weak scale.

In Fig. \ref{mgmrun}, we show the evolution of the various SSB
parameters of the MSSM, starting with the MGM boundary conditions. Here,
the unified gaugino mass is taken to be 300~GeV at $Q=M_{GUT}$. The
compactification scale is taken to be $M_c=10^{18}$~GeV, and other
parameters are fixed to be the same as in Fig.~\ref{mpmg1}. We see that
RG evolution results in GUT scale scalar masses and $A$-parameters that are
substantial fractions of $m_{1/2}$; {\it i.e.} although we have no-scale
\cite{noscale} boundary conditions at the scale $M_c$, there are
substantial deviations from these at $M_{GUT}$. While the
inter-generation splitting is small, the splittings between the {\bf 5}
and the {\bf 10} dimensional matter multiplets, as well as between these
and the Higgs multiplets is substantial.

In Fig.~\ref{mgmmhalf}, we show the variation of several SSB masses at
the scale $Q=M_{GUT}$ with the unified gaugino mass $m_{1/2}$ for the
same values of other parameters as in the previous figure. These masses
then serve as inputs for ISAJET. We note that if $m_{1/2}$ is too small,
the no-scale like boundary conditions lead to incorrect electroweak
symmetry breaking or $m_{\ttau_1} < m_{\tz_1}$. For instance, if
$\tan\beta = 35$ (this allows unification of the $b$ and $\tau$ Yukawa
couplings) with other parameters as in Fig.~\ref{mgmmhalf}, only values
of $m_{1/2}$ larger than 275~GeV are phenomenologically acceptable.

In Table \ref{mgmtable} we show a sample spectrum for this model. We choose
$m_{1/2} =300$~GeV, $\tan\beta=35$ and other parameters as in
Fig.~\ref{mgmmhalf}. The spectrum is not unlike that in the mSUGRA
framework with small $m_0$ so that sleptons are relatively light and
squarks are lighter than the gluino. The chargino and $\tz_2$ almost
exclusively decay via $\tw_1 \to \ttau_1\nu_{\tau}$ and $\tz_2 \to
\ttau_1\tau$, respectively, so that cascade decays of gluinos and squarks
will lead to multi-jet plus multi-tau events, with (soft) leptons as
daughters of the tau. Except for $h$, this scenario is probably beyond
the reach of the Tevatron, but it should be straightforward to study
$\tell_R$ and $\ttau_1$, and probably also detect $\tw_1$ and $\tnu$, at
the NLC. At the LHC a variety of signals should be present.

We have also examined how the mass spectrum changes with variation of the
superpotential couplings $\lambda$ and $\lambda'$. These couplings
cannot be too large in order that they remain perturbative up to
$M_c$. For variation in this range, we found that $m_{10}(GUT)$ and
$m_5(GUT)$ were insensitive to the choice of these couplings, while the
GUT scale values of $m_{{\cal H}_1}$ and $m_{{\cal H}_2}$ as well as
$A_t$ and $A_b$ vary by about 20\% over the entire range of $\lambda$
and $\lambda'$ that we examined. The weak scale spectrum and 
the $\mu$ value are, however, insensitive to the choice of these
parameters; this is presumably because $m_{1/2}$ is significantly larger
than the scalar masses at the GUT scale, so that RG evolution 
between the GUT and
weak scales, rather than from $m_0$, makes the bulk of the
contribution to scalar masses.

\section{Models with non-universal soft terms due to 4--D superstring 
dynamics}

Soft supersymmetry breaking terms obtained from $N=1$ four--dimensional
superstrings, in general, exhibit non-universality at the string scale
\cite{dilaton,ibanez1,ibanez,quevedo}, a notable exception being when
the dilaton is the dominant source of SUSY breaking.  The soft
supersymmetry breaking terms are determined by the K\"ahler potential
$K$ and the gauge kinetic functions $f_a$ of the effective supergravity
theory obtained from the string.  The K\"ahler potential depends on the
hidden sector fields, the dilaton $S$ and the moduli $T$ (there could be
several), and the observable sector fields $C_i$, and it has the form,
\begin{equation}
K=-\log(S+S^*)+K_0(T,T^*)+\widetilde K_{ij}(T,T^*)C_iC^{*j}
\label{KahlerPot}\,.
\end{equation}
To avoid potential problems with FCNCs, we will assume that $\widetilde
K_{ij}= \widetilde{K_i}\delta_{ij}$. 
In addition, the gauge kinetic function in any 4--dimensional superstring
is given at tree level by
\begin{equation}
f_a=k_aS
\label{GaugeKinFun}
\end{equation}
where $k_a$ is the Kac--Moody level of the gauge factor $G_a$, with the 
entire group given by $G=\Pi_aG_a$. The Kac--Moody levels are usually 
taken $k_3=k_2={\textstyle{3\over5}}k_1=1$. Beyond the tree level, $f_a$
would in general also contain a dependence on the moduli fields.  

Supersymmetry is broken when the auxiliary $F$-terms of the hidden sector 
fields acquire vacuum expectation values ($vev$). A convenient way to 
parametrize the vevs (in the case of one modulus) is as follows
\begin{eqnarray}
F^S&=&\sqrt{3}Cm_{3/2}K_{S\bar{S}}^{-1/2}\sin\theta e^{-i\gamma_S}
\nonumber\\
F^T&=&\sqrt{3}Cm_{3/2}K_{T\bar{T}}^{-1/2}\cos\theta e^{-i\gamma_T}
\label{FSFT}
\end{eqnarray}
where $C$ is a constant defined by $C^2=1+V_0/3m_{3/2}^2$, $V_0$ is the
cosmological constant (the vev of the scalar potential), and $m_{3/2}$
is the gravitino mass. Here, $\sin\theta$ is the overlap between the
goldstino and the fermionic component of the dilaton field. Therefore,
$\sin\theta=1$ in the limit where the SUSY breaking is completely due to
the dilaton: {\it i.e.} $\langle F_S \rangle$ is the only relevant $vev$.  
The matrix $K_{n\bar{m}}\equiv\partial_n\partial_{\bar{m}}K$ is called 
the K\"ahler metric and $\gamma_S$ and $\gamma_T$ are possible complex 
phases. 

The soft masses for scalar particles are determined by the K\"ahler
potential in Eq.~(\ref{KahlerPot}) and are given by \cite{ibanez1}
\begin{equation}
m_i^2 =2m_{3/2}^2(C^2-1) +m_{3/2}^2C^2(1+N_i\cos^2\theta),
\label{scalarmass}
\end{equation}
with 
\begin{displaymath}
N_i={{-3(\log{\widetilde K}_i)_{T\bar{T}}} \over (K_0)_{T\bar{T}}}.
\end{displaymath}
We readily see that we can obtain non-universal scalar masses if
$\cos\theta$ is different from zero. We mention that here we have for
simplicity assumed
that there is just one modulus field: multiple moduli are treated in
Ref. \cite{ibanez}.

The gaugino masses are given by
\begin{equation}
M_a=\half({\mathrm{Re}}f_a)^{-1}F^m\partial_mf_a
=\sqrt{3}Cm_{3/2}
({\mathrm{Re}}f_a)^{-1}k_a{\mathrm{Re}}Se^{-i\gamma_S}\sin\theta,
\label{GaugiMass}
\end{equation}
where the gauge coupling constants are ${\mathrm{Re}}f_a=1/g_a^2$.
In the last equality, we have used the fact that (at tree level) the
gauge kinetic function in Eq.~(\ref{GaugeKinFun}) depends only on the
dilaton field $S$, so that the tree level gaugino masses are independent
of the moduli sector. Model-dependent corrections to this may, however,
be significant, particularly when dilaton contributions to SUSY breaking
are small. 

Expressions for $A$-parameters may also be found in
Ref. \cite{ibanez1}. These depend on additional parameters, and
generically also on the unknown phases $\gamma_S$ and $\gamma_T$ (as
well as on additional direction cosines in the multi-moduli case). For the
single modulus case, the
form of $A$ is given by,\footnote{We have flipped the sign of $A$ to
conform to our convention where the soft trilinear term is written as 
$A_{ijk}f_{ijk}{\tilde{C}}_i{\tilde {C}}_j{\tilde{C}}_k$ in the Lagrangian and
not the scalar potential, with $f_{ijk}$ being the corresponding
superpotential coupling.}
\begin{equation}
A_{ijk} = \sqrt{3}m_{3/2}C(e^{-i\gamma_S}\sin\theta 
+e^{-i\gamma_T}\omega_{ijk}(T,T^*)\cos\theta),
\label{Aeqn}
\end{equation}
where $\omega_{ijk}$ depend on the K\"ahler and superpotentials. 
Fortunately, in
many cases of interest, these model-dependent parameters either vanish
or assume a simple form.

We should mention that these expressions for the soft-SUSY breaking
masses and $A$-parameters are valid for these parameters renormalized at
the string scale. As always, these have then to be evolved down to the
weak scale for use in phenomenological analysis.  We now consider some
special cases to illustrate the forms of (string scale) non-universality
that may occur in this general framework.

\subsection{Large--T limit of Calabi--Yau compactifications}

Because of the complexity of the world--sheet instanton  and sigma model
contributions, the general form of
the K\"ahler potential of generic Calabi--Yau $(2,2)$ compactifications
is not known. The gauge group is $E_6\times E_8$, with matter in the {\bf 27}
dimensional representation of $E_6$.
It is usual to analyze the large $T$ (in practice $2-3<|T|<20-30$,
large enough so that world sheet instanton contributions can be
neglected, but not so large that string threshold corrections invalidate
perturbation theory) limit of these theories. In this limit the K\"ahler
potential takes a simple form \cite{ibanez1}:
\begin{equation}
K=-\log(S+S^*)-3\log(T+T^*)+\sum_i{{|C_i|^2}\over{T+T^*}}\,,
\label{KahlerCY}
\end{equation}
and the gauge kinetic function is given by Eq.~(\ref{GaugeKinFun}) at tree 
level. In this case the gaugino mass is
\begin{equation}
m_{1/2}=\sqrt{3}Cm_{3/2}
%{{k_a{\mathrm{Re}}S}\over{{\mathrm{Re}}f_a}}
\sin\theta e^{-i\gamma_S},
\label{GauCYlargT}
\end{equation}
while Eq.~(\ref{scalarmass}) for the scalar masses reduces to,
\begin{equation}
m_0^2=m_{3/2}^2C^2\sin^2\theta+2m_{3/2}^2(C^2-1)
\label{scalCYlargT}
\end{equation}
which simplifies even further if the cosmological constant vanishes
($C=1$). Notice that we find universality of
soft scalar masses, even though we are not in the dilaton dominated SUSY
breaking scenario.

In the $C=1$ case, we see that $|m_{1/2}|=\sqrt{3}m_0$, so that the
gaugino mass always exceeds the scalar mass at the string scale. This
relation obviously puts a significant constraint on SUSY
phenomenology. Since this is a special case of the mSUGRA scenario whose
phenomenological implications have been discussed at length in the
literature, we will not mention this any further. 

There are, however, arguments in the literature \cite{coleman} that
suggest that the observed cosmological constant (which is bounded to be
smaller than $\sim (3 \ meV)^4$) may not be directly connected to $V_0$;
then, $C$ could differ from unity, and the gaugino mass may (depending
on the value of $C$ and the goldstino angle $\theta$) be even smaller
than $m_0$, but for an appreciable effect, $C-1$ would have to deviate
by many orders of magnitude\footnote{It should be appreciated that even $C=1.1$
is an enormous value relative to the bound $C-1 \alt 10^{-87}$
that we would get if we took $V_0$ to be related to the observed value
of $\Lambda$.} from the bound that would have resulted assuming $V_0$
was the observed cosmological constant. 

Finally, in this limit, the parameters $\omega_{ijk}$ in Eq.~(\ref{Aeqn})
vanish so that  
\begin{displaymath}
A_{ijk} = \sqrt{3}m_{3/2}Ce^{-i\gamma_S}\sin\theta.
\end{displaymath}

In the single modulus large $T$ case that we have been discussing, effects of
the sigma--model loop contribution and the non--perturbative instanton
contribution to the K\"ahler potential are known \cite{KM}.
We still obtain universality of soft SUSY breaking parameters, with
gaugino masses given by Eq.~(\ref{GauCYlargT}) and scalar masses and the
$A$ parameter (in the case $C=1$) modified to,
\begin{equation}
m_0^2=m_{3/2}^2\left[1-\cos^2\theta\left(1-\Delta(T,T^*)\right)\right],
\label{m0CYgen1mod}
\end{equation}
and 
\begin{equation}
A=\sqrt{3}m_{3/2}\left [e^{-i\gamma_S}\sin\theta + 
\omega(T,T^*)e^{-i\gamma_T}\cos\theta \right ].
\label{A0CYgen1mod}
\end{equation}
Here $\Delta$ and $\omega$ corresponds to the sigma--model and instanton
contributions (the latter are negligible):
the numerical values of these are model dependent, but
$\Delta\approx 0.4$ and $\omega=0.17$ have been quoted \cite{KM} for a
typical model. Notice that although these corrections do not
lead to non-universality, we lose the earlier prediction
$m_{1/2}=\sqrt{3}m_0$: now, the soft scalar mass may even exceed the
corresponding gaugino mass if $\cos^2\theta$ is sufficiently large.

\subsection{General Calabi--Yau compactifications}

There is no reason to believe that there is just a single modulus field $T$.
In the multi--moduli case the parametrization of the $vevs$ of the moduli in 
Eq.~(\ref{FSFT}) is modified to \cite{ibanez},
\begin{equation}
F^{T_i}=\sqrt{3}Cm_{3/2}K_{T_i\bar{T}_i}^{-1/2}\cos\theta\Theta_i 
e^{-i\gamma_{T_i}},
\end{equation}
where we have assumed the K\"ahler metric to be diagonal to avoid any
FCNC problems. Here $\Theta_i$ are direction cosines that parametrize
the direction of the $vev$ in moduli space. Indeed the more general case
of an off-diagonal metric has also been examined in Ref.~\cite{KM} where
a more general parametrization of the $vevs$ of the moduli may be
found. In this general case, the scalar masses are non-diagonal and the
mass squared matrix assumes the form,
\begin{equation}
m_{ij}^2=m_{3/2}^2\left[\delta_{ij}-\cos^2\theta\left(\delta_{ij}
-\Delta_{ij}(T_k,T^*_k)\right)\right]
\label{mijCYmulMod}
\end{equation}
where $\Delta_{ij}$ depends on the moduli and on the direction of the
$vev$ in the moduli space. Notice that the model-dependent $\Delta_{ij}$
would be strongly constrained by experimental data on flavour mixing. We
are, however, not aware of a realistic model in which such constraints
may be analyzed. We also note that the presence of a (even diagonal)
matrix $\Delta$ in Eq.~(\ref{mijCYmulMod}) would be a source of
non-universality of scalar masses.

\subsection{Orbifold models with large threshold corrections}

An example of such a model is the so-called $O$-$I$ model discussed by
Brignole {\it et al.} \cite{ibanez1}. In orbifold 
compactifications the coefficient $\widetilde K_{ij}$ which determines
the soft masses has the form $(T+T^*)^{n_i}$, where $n_i$ is the modular 
weight of the matter field $C_i$. The K\"ahler potential is in this case:
\begin{equation}
K=-\log(S+S^*)-3\log(T+T^*)+\sum_i|C_i|^2(T+T^*)^{n_i}
\label{KahlerOrbi}
\end{equation}
Gauge unification in good agreement with low energy data is achieved by
assigning the following modular weights for the massless fields:
$n_{Q}=n_{D}=-1$, $n_{U}=-2$, $n_{L}=n_{E}=-3$, and
$n_{H_d}+n_{H_u}=-5$ or $-4$, together with a large value for the
modulus field, ${\mathrm T}\approx16$, which then results in large
threshold corrections. Under these conditions the gaugino masses are
non-universal at the string scale:
\begin{eqnarray}
M_1&=&1.18\sqrt{3}m_{3/2}\left[\sin\theta+
2.9\times10^{-2}(B'_1/k_1)\cos\theta\right]
\nonumber\\
M_2&=&1.06\sqrt{3}m_{3/2}\left[\sin\theta+
2.9\times10^{-2}(B'_2/k_2)\cos\theta\right]
\label{gaugmassOI}\\
M_3&=&1.00\sqrt{3}m_{3/2}\left[\sin\theta+
2.9\times10^{-2}(B'_3/k_3)\cos\theta\right]
\nonumber
\end{eqnarray}
where $B'_a\equiv b'_a-k_a\delta_{GS}$ are given by
$B'_1=-18-k_1\delta_{GS}$, $B'_2=-8-k_2\delta_{GS}$, and
$B'_3=-6-k_3\delta_{GS}$ if $n_{H_d}+n_{H_u}=-5$, and $k_a$ as
specified previously. Here, the parameter $\delta_{GS}$ is a model
dependent negative integer and $m_{3/2}$ and $\theta$ are the gravitino
mass and the goldstino angle as before.  To obtain
Eqs.~(\ref{gaugmassOI}), it is assumed\cite{ibanez1} that string
threshold corrections lead to an apparent unification of the couplings
at the ``$GUT$ scale'' rather than at the string scale. Of course, since
there is no $GUT$ these couplings continue to evolve and diverge when
evolved from the ``$GUT$ scale'' to the one order of magnitude larger string
scale. The coefficients in front of the gaugino mass formulae reflect
just this difference in the gauge couplings at the string scale. In
other words, if $\sin\theta=1$, gaugino masses (while slightly different
at the string scale) would be universal at $Q=M_{GUT}$: non-universality
of GUT scale gaugino masses occurs only due to the loop correction
proportional to $\cos\theta$ in Eqs.~(\ref{gaugmassOI}).  Finally, we
note that if $n_{H_d}+n_{H_u}=-4$, the gaugino masses are
obtained from Eqs.~(\ref{gaugmassOI}) by modifying the coefficients
$B'_i$ to $B'_1=-17-k_1\delta_{GS}$ and $B'_2=-7-k_2\delta_{GS}$ while
$B'_3$ does not change.

The string scale scalar masses and $A$ parameters
depend on the modular weights, and (assuming zero
cosmological constant) are given by,
\begin{eqnarray}
m^2_Q=m^2_D&=&m_{3/2}^2\left[1-(1-\delta_{GS}\times10^{-3})^{-1}
\cos^2\theta\right]\,,
\nonumber\\
m^2_U&=&m_{3/2}^2\left[1-2(1-\delta_{GS}\times10^{-3})^{-1}
\cos^2\theta\right]\,,
\label{OIscalarmass}\\
m^2_L=m^2_E&=&m_{3/2}^2\left[1-3(1-\delta_{GS}\times10^{-3})^{-1}
\cos^2\theta\right]\,,
\nonumber
\end{eqnarray}
and
\begin{equation}
A_{ijk}=\sqrt{3}m_{3/2}\sin\theta \pm 
m_{3/2}\cos\theta (1-\delta_{GS}\times10^{-3})^{-1/2} 
(3+n_i+n_j+n_k),
\label{OIA}
\end{equation}
where the terms with $\delta_{GS}$ come from radiative corrections, and
the sign ambiguity reflects the possible relative phase between
$\gamma_S$ and $\gamma_T$ (we take the $A$-parameters to be real).  Note
that if $\sin\theta=1$, the scalar masses and $A$-parameters are
universal at the string scale: RG evolution would then introduce a small
non-universality at $M_{GUT}$.

In Fig.~\ref{oimass} we plot different soft masses at the string scale
as a function of $\sin\theta$ in the $O$-$I$ model. There is a sign
ambiguity since $\cos\theta$ could be negative. We have chosen
$\cos\theta > 0$
and fixed
$m_{3/2}=200$ GeV, $\delta_{GS}=0$, and, for the evaluation of gaugino masses,
$n_{H_u}+n_{H_d}=-5$. We set the phases $\gamma_S$ and $\gamma_T$
to be zero. Scalar masses are universal in the dilaton
dominated scenario and radiative corrections do not spoil this
universality. On the contrary, gaugino masses are not universal at
$\sin\theta=1$, but as explained above, there is (approximate)
universality at $M_{GUT}$. Values of $\cos\theta\gsim 1/\sqrt{3}$
($\sin\theta\lsim 0.8$) yield negative slepton soft squared masses and
may be unacceptable;\footnote{It may be
possible to have these squared masses negative at a high scale as long as
they are positive near the weak scale.} hence
the dilaton field is necessarily the most important source of SUSY
breaking. Except close to the lowest acceptable values of $\theta$,
deviations from universality in the scalar sector are thus limited.

To facilitate simulation of such a scenario, we have introduced 
into ISAJET versions $\geq 7.50$ the
``SUSY Boundary Condition Scale'' ($SSBCSC$ keyword) 
option into ISAJET that allows the
user to input a chosen scale $Q_{max}$ up to which the MSSM is assumed to be
valid. 
The values of SUSY breaking masses and $A$-parameters of the MSSM
as given by any theory valid at the scale beyond $Q_{max}$ would then be
used as inputs to ISAJET, which would then evolve them down to the weak
scale and generate SUSY events as usual. For the case at hand, $Q_{max}$
would be the string scale, and the gaugino masses, scalar masses and
$A$-parameters as given by Eqs.~(\ref{gaugmassOI}) - (\ref{OIA}), the
boundary conditions for the RGE. We stress, however, that $Q_{max}$ need
not be larger than $M_{GUT}$. For instance, in $SO(10)$ models,
$Q_{max}$ would be the mass of the right-handed neutrino, or in $E_6$
models, the mass scale where the additional particles in the {\bf 27}
dimensional representation and any extra $Z'$ bosons all decouple, leaving
the MSSM spectrum.

We give an example of the SUSY spectrum in the O-I scenario in Table
\ref{o-scenarios}.  In this example, we have fixed $\tan\beta=4$,
$\sin\theta = 0.85$ (with $\cos\theta>0$) and have taken $n_{H_u}= -3$,
with other parameters as in Fig.~\ref{oimass}.  Since the value of $B$
depends on how $\mu$ is generated, we have treated $\tan\beta$ as a free
parameter, and eliminated $B$ in its favour, using the constraints
given by radiative electroweak symmetry breaking.  We fix the string
scale to be $4\times 10^{17}$~GeV. Despite the fact that string
scale slepton masses are considerably smaller than those of squarks (see
Fig.~\ref{oimass}), the spectrum is qualitatively very similar to that
in the mSUGRA framework with $m_{\tq} \sim m_{\tg}$.

\subsection{Orbifold models with small threshold corrections}

In the $O$-$I$ model, $\sin\theta$ was restricted to be large, so that the
parameters of phenomenological interest were qualitatively similar to
the mSUGRA scenario. To allow a wider range of $\sin\theta$ we consider
a model where all the modular weights are $-1$. As noted in Ref. \cite{ibanez1}
string threshold corrections cannot account for gauge coupling
unification, which has then to be attributed to some different physics.
Unlike the $O$-$I$ model where a large value of ${\mathrm Re}T$ was
needed to accommodate coupling constant unification, we will, following
Brignole {\it et al.} \cite{ibanez1} use  
${\mathrm Re}T\approx1.2$ and refer to this 
as the $O$-$II$ model. As before,
the gaugino masses are non-degenerate at the string scale (again, for
$\sin\theta=1$, these would be universal at ``$M_{GUT}$'') and given by:
\begin{eqnarray}
M_1&=&1.18\sqrt{3}m_{3/2}\left[\sin\theta+
4.6\times10^{-4}(B''_1/k_1)\cos\theta\right]
\nonumber\\
M_2&=&1.06\sqrt{3}m_{3/2}\left[\sin\theta+
4.6\times10^{-4}(B''_2/k_2)\cos\theta\right]
\label{gaugmassOII}\\
M_3&=&1.00\sqrt{3}m_{3/2}\left[\sin\theta+
4.6\times10^{-4}(B''_3/k_3)\cos\theta\right]
\nonumber
\end{eqnarray}
with $B''_1=11-k_1\delta_{GS}$, $B''_2=1-k_2\delta_{GS}$, and
$B''_3=-3-k_3\delta_{GS}$. On the other hand, the scalar masses ($V_0=0$)
and $A$ parameters are all degenerate and equal to
\begin{equation}
m^2_Q=m^2_D=m^2_U=m^2_L=m^2_E=m_{3/2}^2\left[1-
(1-\delta_{GS}\times10^{-3})^{-1}\cos^2\theta\right]\,,
\label{OIIscalarmass}
\end{equation}
and
\begin{equation}
A_{ijk}= \sqrt{3}m_{3/2}\sin\theta,
\end{equation}
at the string scale.

If $\sin\theta \sim 1$ the spectra should be the same as in the $O$-$I$
model discussed previously. For smaller values of $\sin\theta$, the
degeneracy in the string scale scalar masses still remains. The most
important difference between the two scenarios is that very small values
of $\sin\theta$ are now permitted; {\it i.e.} the dilaton contribution
need not necessarily dominate SUSY breaking. If $\sin\theta$ is very
small so that the $\cos\theta$ terms are the dominant contributions to
the gaugino mass, we see that (depending on the value of $\delta_{GS}$)
the GUT scale gluino mass may be much smaller than the corresponding
electroweak gaugino masses. Indeed it is possible~\cite{gunion} to arrange 
scenarios where the gluino is the 
LSP\cite{bcg}. 
The additional parameters also allow the possibility $M_1 \simeq
M_2$ so that the lighter chargino and the two lighter neutralinos (and
sometimes also the gluino) are
all very degenerate. Such scenarios pose interesting experimental challenges
\cite{gunion}.

In Fig.~\ref{oiimass} we illustrate the gaugino and scalar soft masses
at the string scale as a function of $\sin\theta$ in the $O$-$II$
model. Again, we take $m_{3/2}=200$~GeV, $\cos\theta>0$, and ignore all
phases. We choose $\delta_{GS}=-5$. The masses decrease as $\sin\theta$
decrease but they do not vanish at $\sin\theta=0$ due to one--loop
effects. Of course, for very small values of $\sin\theta$
phenomenological considerations require $m_{3/2}$ to be significantly
larger.  In the extreme case of moduli-dominated SUSY breaking, gaugino
masses can be smaller than scalar masses, but generally speaking scalar
masses are smaller than gaugino masses at the unification scale.

In the last three columns of Table \ref{o-scenarios} we illustrate three
examples of $O$-$II$ model spectra. Again, we fix $\tan\beta=4$, $\mu>0$
and take $\delta_{GS} =-5$, to be in the region which can potentially
yield~\cite{gunion} roughly equal masses for all the MSSM
gauginos. First, we choose an $O$-$II$ scenario with parameters close to
those of the $O$-$I$ model in the previous column: $m_{3/2}=200$~GeV and
$\sin\theta=0.85$. This is the ``typical'' case for such a model. In
this case, the $\sin\theta$ terms in Eq.~(\ref{gaugmassOII}) completely
dominate, and the resulting spectrum is again very similar to that in
the mSUGRA framework (with $m_{\tq} \sim m_{\tg}$).

In the next column, we show a spectrum for the case $\sin\theta=0$, the
extreme case\footnote{It does not matter whether we take $\theta=0$ or
$\theta=\pi$ since the sign of the gaugino mass has no import for
physics.} of moduli-dominated SUSY breaking. Here, because of the small
coefficient $4.6\times 10^{-4}$ in the expressions for gaugino masses,
we have to choose $m_{3/2}$ to be large.  We fix $m_{3/2}=60$~TeV. For
this case, we have taken $m_t=180$~GeV, since we found that electroweak
symmetry was not broken\footnote{The scalars start at a very large mass
at the string scale, and the top Yukawa is not large enough to drive a
Higgs mass squared eigenvalue negative at the scale $Q
=\sqrt{m_{\tst_L}m_{\tst_R}}$ where the effective potential is evaluated
in ISAJET~\cite{isajet}. We should mention that this is sensitive to the
top mass radiative corrections that have been included~\cite{isajet} in
ISAJET versions $\geq 7.48$. These radiative corrections decrease the
top Yukawa coupling by a few percent, and in this case, this is just
sufficient to preclude electroweak symmetry breaking.} for
$m_t=175$~GeV.  Since the (common) string-scale scalar mass is much
bigger than the corresponding gaugino masses, the scalars are all
roughly degenerate, and their spectrum is close to that of the
corresponding mSUGRA spectrum with $m_{\tq} \sim m_{\tell} \gg m_{\tg}$
({\it i.e.} $m_0 \gg m_{1/2}$).  The gluinos, charginos and neutralino
spectrum is quite different from that in the mSUGRA model: even though
the lighter chargino and neutralinos are gaugino-like,
$m_{\tw_1}=m_{\tz_2}=m_{\tz_1} = 0.7m_{\tg}$. This is because by
choosing $\delta_{GS}$ we can adjust $M_1: M_2: M_3$ at the string
scale.  By a careful adjustment of parameters the gluino mass can even
be brought closer to the chargino and neutralino masses.  Experiments at
the Fermilab Tevatron may be sensitive to this scenario.

To emphasize that the novel scenarios shown in Ref.~\cite{gunion}
obtain only for a very limited range of parameters, in the last column
we show the spectrum for $\sin\theta=0.005$ (with $\cos\theta>0$), with
all other parameters (including $m_t$) as for the $\sin\theta=0$ case. We
see that even for this tiny value of $\sin\theta$, the $\sin\theta$
terms in Eq.~(\ref{gaugmassOII}) are comparable to (or even dominate) the
$\cos\theta$ terms, and the spectrum is qualitatively different. While
the sfermions are once again extremely heavy, the gluino, chargino and
neutralino masses are now approximately as in the mSUGRA
framework. Sparticle detection in this scenario would only be possible
at the LHC. Our purpose in showing this (possibly unacceptably heavy)
spectrum is only to emphasize the qualitative difference from the
$\sin\theta=0$ case. Of course, if $m_{3/2}$ is chosen to be 15~TeV,
many more sparticles would be in the accessible range, but
the spectrum would then be much like the canonical mSUGRA case with large $m_0$.
  
\section{Models with non-universal soft terms due to M--theory dynamics}

It was proposed that M--theory, \ie, an 11--dimensional supergravity on
a manifold where two $E_8$ gauge multiplets are restricted to the two
10--dimensional boundaries, is equivalent to the strong coupling limit
of $E_8\times E_8$ heterotic string theory \cite{HW}. 
It may be argued that
M--theory is a better candidate than the weakly coupled string to
explain low energy physics and unification. 
After compactifying the
11--dimensional M--theory, a 4--dimensional effective theory emerges
which can reconcile the reduced Planck scale
$M_P\approx2.4\times10^{18}$ GeV, the grand unification scale
$M_{GUT}\approx3\times10^{16}$ GeV, and $\alpha_{GUT}$, in a way that
the weakly coupled heterotic string theory cannot. An interesting
feature of the 4--dimensional effective SUGRA is that, in first
approximation, the gauge kinetic function, the superpotential, and the
K\"ahler potential do not change when moving from the weakly coupling
heterotic string case to the M--theory case by changing the value of
%$\alpha(T+T^*)$, where $\alpha$ is an integer and $T$ is 
the modulus field.

\subsection{One modulus case}

Supersymmetry is broken when the auxiliary components of the dilaton
field $S$ and the modulus field $T$ aquire non--zero $vevs$, as
discussed in the last Section. The low energy effective supergravity
theory~\cite{CKM} obtained from a specific Calabi-Yau compactification
in M--theory is a Yang-Mills gauge theory with $E_6$ as the gauge group.
The gauge kinetic function is given by,
\begin{equation}
f_{E_6}=S+\alpha T
\label{GauKinFunMth}
\end{equation}
where $\alpha$ is an integer while the corresponding K\"ahler potential is
\begin{equation}
K=-\log(S+S^*)-3\log(T+T^*)+\left[{3\over{T+T^*}}+{{\alpha}\over{S+S^*}}
\right]\sum_i|C_i|^2\,,
\label{KahlerMth}
\end{equation}
where $C_i$ again denote the observable fields. Adopting the same
parametrization as in Eq.~(\ref{FSFT}) above, we find that with the
K\"ahler potential of Eq.~(\ref{KahlerMth}) the soft SUSY breaking
parameters are universal and given by,
\begin{equation}
m_{1/2}={{\sqrt{3}Cm_{3/2}}\over{1+x}}\left[\sin\theta e^{-i\gamma_S}+
{x\over{\sqrt{3}}}\cos\theta e^{-i\gamma_T}\right],
\label{MhalfMtheory}
\end{equation}
\begin{eqnarray}
m_0^2&=&m_{3/2}^2(3C^2-2)-{{3C^2m_{3/2}^2}\over{(3+x)^2}}\bigg[
x(6+x)\sin^2\theta+(3+2x)\cos^2\theta
\nonumber\\&&\qquad\qquad\qquad\qquad\qquad\qquad
-2\sqrt{3}\,x\sin\theta\cos\theta\cos(\gamma_S-\gamma_T)\bigg],
\label{m0Mtheory}
\end{eqnarray}
and
\begin{equation}
A={{\sqrt{3}Cm_{3/2}}\over {3+x}}\left [(3-x)\sin\theta e^{-i\gamma_S} 
  +\sqrt{3}x\cos\theta e^{-i\gamma_T} \right ],
\label{AMtheory}
\end{equation}
where,
\begin{equation}
x\equiv {{\alpha(T+T^*)}\over{S+S^*}}\,,
\label{xdef}
\end{equation}
The range of $x$ is $0\le x\le 1$.

Note that in the weak coupling limit $x\rightarrow 0$, we recover from
Eqs.~(\ref{MhalfMtheory}), (\ref{m0Mtheory}) and (\ref{AMtheory}) the
gaugino and scalar masses as well as the $A$-parameter in the large
$T$--limit of Calabi--Yau compactifications in Eqs.~(\ref{GauCYlargT})
and (\ref{scalCYlargT}) respectively.

In Fig.~\ref{m1mod} we show the dependence on the goldstino angle of the
universal gaugino and scalar masses, $m_{1/2}$ and $m_0$ respectively.
We consider zero cosmological constant and three values of $x$. The strong 
coupling limit corresponds to $x=1$ and for comparison $x=0.5$ 
and $x=0$ are also plotted. 

We remind the reader that the soft parameters obtained above are for an 
$E_6$ gauge theory. In order to obtain a realistic low energy theory, we
have to know how the symmetry group is reduced to the MSSM gauge group,
which in turn will depend on the details of the theory at the high
scale. It is possible that there may be additional $TeV$ scale
supermultiplets in the particle spectrum, or even extra gauge
bosons\cite{hewett}. Moreover, depending on how $E_6$ breaks to $SU(3)
\times SU(2) \times U(1)$, additional $D$-term contributions (see
Sec.~V) which break the universality of scalar masses may also be present.

\subsection{Multi-moduli case}

As before, the situation in the multi-moduli case can be more
complicated. A toy example with three moduli fields and three observable
fields has been considered in Ref. \cite{Li}. The K\"ahler potential
and gauge kinetic function of the effective theory is written as, 
\begin{equation}
K=-\log(S+S^*)-\sum^3_{j=1}\log(T_j+T_j^*)+\left[
2+{2\over3}\sum_{j=1}^3{{\alpha_j(T_j+T_j^*)}\over{S+S^*}}
\right]\sum_{i=1}^3{{|C_i|^2}\over{T_i+T_i^*}}\,.
\label{KahlerMth3m}
\end{equation}
and 
\begin{equation}
f_a= S+\sum^3_{i=1}\alpha_i T_i
\end{equation}
This then yields a universal mass for the gaugino and a universal
$A$-parameter, but non-universal masses (and no mixing) for the scalars.
While the gaugino and scalar masses as well as the
$A$-parameter depend on the parameters and fields in the K\"ahler
potential, the splitting $\delta m^2$ (between the scalars) appears to
depend only on the orientation of the $vevs$ of the auxiliary components
of the moduli and on the goldstino angle $\theta$. Since our
focus is on sources of non-universality in realistic scenarios that can
potentially be of phenomenological interest, we merely note that
multiple moduli could be a source of non-universality of scalar masses,
but do not exhibit results for this toy model here.

\section{Concluding Remarks}

While weak scale supersymmetry is a well-motivated idea, the physical
principles that fix the multitude of SUSY breaking parameters are not
known. Without any sparticle signals to provide clues, we do
not have any guidance as to what these might be.  The scale of this new physics
may be as low as a few hundred TeV as in models with low energy SUSY
breaking mediated by gauge interactions, or as high as $M_{GUT}-M_P$ as
in frameworks where SUSY breaking is mainly mediated by
gravity. Observable sparticle masses and mixing patterns, and via these
weak scale SUSY phenomenology, are determined by the physics behind SUSY
breaking and how this is communicated to the observable sector.  Turning
this around, measurement of sparticle properties may provide clues about
physics at energy scales that would be inaccessible to experiments in
the foreseeable future.

Most early phenomenological analyses have been done within the framework
of the mSUGRA model or the mSUGRA-motivated MSSM (where {\it ad hoc}
relations between SSB parameters were assumed). In the last few years,
phenomenological aspects of gauge-mediated SUSY breaking have also been
examined in some detail. Both these models rest
upon untested assumptions about physics at high energies. The good thing
is that some of these assumptions will be directly testable if
sparticles are discovered and their properties are measured~\cite{future}.
Nevertheless, it seems worthwhile to look at other viable alternatives
for physics at energy scales much beyond the weak scale, with a view to
see if there are direct ramifications for sparticle signals in future
experiments. A serious study of this would entail SUSY simulation at
colliders in a wide variety of models with features different from the
mSUGRA paradigm, which is characterized by universality of SSB
parameters at a scale $Q \sim M_{GUT}$.

Our study represents a first step in this direction. Here, we have
surveyed a number of proposals for high scale physics that lead to
non-universality of the SSB parameters in the MSSM, which we regard as
the effective theory at a sufficently low mass scale. These range from
relatively minor modifications of the mSUGRA $SU(5)$ GUT model, where,
{\it e.g.}  unification of scalar masses and $A$-parameters is assumed
to occur at $M_P$ (so that RG evolution induces some non-universality at
$Q=M_{GUT}$), to major modifications involving conceptually new ideas
for high scale physics (new hypercolour interactions, string physics) or
the mediation of SUSY breaking to the observable sector (anomaly
mediated SUSY breaking, gaugino mediated SUSY breaking). 
Other proposals that fall somewhat between
these two extremes include models with larger unifying groups that
naturally have additional non-universal contributions to scalar masses,
or models where special boundary conditions on SSB parameters lead to
unusual RG evolution and non-degeneracy of sparticle masses. For each of
these scenarios, we have outlined the underlying physical
ideas, delineated the parameter space in terms of which SUSY
phenomenology might be analyzed, and discussed SUSY event generation
using the simulation program ISAJET~\cite{isajet}. 
A variety of improvements to the ISAJET program have been made to
allow event generation in the models discussed in this paper.
These improvements are characterized by ISAJET keyword inputs,
including $NUSUGi$ for non-universal masses, $SUGRHN$ for models
with a right-handed neutrino contribution, such as $SO(10)$, $SSBCSC$
for user choice as to when the MSSM becomes valid, and $AMSB$ for
anomaly-mediated SUSY breaking models.
Where possible, we present sample
spectra, and allude to the important phenomenological differences from
the reference mSUGRA framework.

A detailed phenomenological analysis of each one of these scenarios is
beyond the scope of the present work. Our hope though is that this study
will facilitate and spur such analyses. Except for unusual cases where
extreme degeneracies between sparticle masses result \cite{gunion}, we
do not expect the reach of various future facilities (expressed in terms
of physical sparticle masses) to qualitatively differ between the
various scenarios. However, a careful examination of these will help us
assess what we can hope to learn about high scale physics if sparticles
are discovered and their properties measured. Careful examination of
physical implications of a variety of viable alternatives for the
underlying theory will also help increase our understanding of the sort
of analyses that might be needed to discriminate between these. In view
of the potential pay-off, we believe that such studies will be very
worthwhile.

%%%%%%%%%%%%%%%%%%%%%%%%% ACKNOWLEDGEMENTS %%%%%%%%%%%%%%%%%%%%%%%%%%%%%%%%%%%%%
%
%\newpage
\acknowledgments
We thank B. de Carlos, M. Drees, L.~Iba\~nez, 
P. Mercadante, C. Mu\~noz, 
and T. Li for helpful communications, and M.~Drees again for valuable comments
on the manuscript.
This research was supported in part by the U.~S. Department of Energy
under contract number DE-FG02-97ER41022 and DE-FG-03-94ER40833. M.A.D.
was also supported in part by CONICYT grant 1000539.
%

%%%%%%%%%%%%%%%%%%%%% REFERENCES %%%%%%%%%%%%%%%%%%%%%%%%%%%%%%%%%%%%%%%%%%%%%%
%

\newpage
%
%%%%%%%%%%%%%%%%%%%%%%%%%% TABLES %%%%%%%%%%%%%%%%%%%%%%%%%%%%%%%%%%%%%%%%%%%
%
% Stupid RevTeX.
\iftightenlines\else\newpage\fi
\iftightenlines\global\firstfigfalse\fi
\def\dofig#1#2{\epsfxsize=#1\centerline{\epsfbox{#2}}}

\begin{table}
\begin{center}
\caption{Input and output parameters for an $SU(5)$ case study}
\bigskip
\begin{tabular}{lcc}
\hline
parameter & scale & value\\
\hline
$m_0$ & $M_P$ & 150 \\
$m_{1/2} $ & $M_{GUT}$ & 200 \\
$A_0$ & $M_P$ & 0 \\
$\tan\beta$ & $M_{weak}$ & 35 \\
$\mu $ & $M_{weak}$ & $<0$ \\
$g_{GUT}$ & $M_{GUT}$ & 0.717  \\
$f_t$ & $M_{GUT}$ & 0.534  \\
$f_b=f_\tau$ & $M_{GUT}$ & 0.271 \\
$\lambda$ & $M_{GUT}$ & 1 \\
$\lambda'$ & $M_{GUT}$ & 0.1 \\
$m_{10}^{1,2}$ & $M_{GUT}$ & 194.4 \\
$m_{5}^{1,2}$ & $M_{GUT}$ & 180.8 \\
$m_{10}^{3}$ & $M_{GUT}$ & 183.8 \\
$m_{5}^{3}$ & $M_{GUT}$ & 177.7 \\
$m_{H_d}$ & $M_{GUT}$ & 107.8 \\
$m_{H_u}$ & $M_{GUT}$ & 96.2 \\
$A_t$ & $M_{GUT}$ & -87.6 \\
$A_b=A_\tau$ & $M_{GUT}$ & -77.6 \\
\hline
\label{tsu5_1}
\end{tabular}
\end{center}
\end{table}
\begin{table}
\begin{center}
\caption{Weak scale sparticle masses and parameters (GeV) for mSUGRA 
and for an $SU(5)$ case study.}
\bigskip
\begin{tabular}{lcc}
\hline
parameter & mSUGRA & $SU(5)$ \\
\hline
$m_{\tg}$ & 512.0 & 515.0 \\
$m_{\tu_L}$ & 468.0 & 484.0 \\
$m_{\td_R}$ & 454.7 & 463.2 \\
$m_{\tst_1}$ & 335.6 & 337.8 \\
$m_{\tb_1}$ & 375.4 & 375.2 \\
$m_{\tell_L}$ & 212.8 & 235.4 \\
$m_{\tell_R}$ & 174.5 & 213.8 \\
$m_{\ttau_1}$ & 124.3 & 151.1 \\
$m_{\tw_1}$ & 150.1 & 155.3  \\
$m_{\tz_2}$ & 150.3 & 155.3  \\
$m_{\tz_1}$ & 80.8 &  81.5 \\
$m_h$ & 111.0 & 111.6 \\
$m_A$ & 210.4 & 216.4 \\
$\mu$ & -263.8 & -304.3 \\
\hline
\label{tsu5_2}
\end{tabular}
\end{center}
\end{table}
\begin{table}
\begin{center}
\begin{small}
\begin{tabular}{|c|ccc|ccc|}
\hline
\ & \multicolumn{3}{c|} {$\mgut$} & \multicolumn{3}{c|}{$\mz$} \cr
$F_{\Phi}$
& $M_3$ & $M_2$ & $M_1$
& $M_3$ & $M_2$ & $M_1$ \cr
\hline
${\bf 1}$   & $1$ &$\;\; 1$  &$\;\;1$   & $\sim \;6$ & $\sim \;\;2$ &
$\sim \;\;1$ \cr
${\bf 24}$  & $2$ &$-3$      & $-1$  & $\sim 12$ & $\sim -6$ &
$\sim -1$ \cr
 ${\bf 75}$  & $1$ & $\;\;3$  &$-5$      & $\sim \;6$ & $\sim \;\;6$ &
$\sim -5$ \cr
${\bf 200}$ & $1$ & $\;\; 2$ & $\;10$   & $\sim \;6$ & $\sim \;\;4$ &
$\sim \;10$ \cr
\hline
\end{tabular}
\end{small}
\smallskip
\caption{Relative gaugino masses at $\mgut$ and $\mz$
in the four possible $F_{\Phi}$ irreducible representations.}
\label{masses}
\end{center}
\end{table}
\begin{table}
\begin{center}
\caption{Weak scale sparticle masses and parameters (GeV) for the 
four cases of singlet and non-singlet hidden sector vevs in $SU(5)$.
For each case, we take $(m_0,\ M_3^0,\ A_0)= (100,\ 150,\ 0$ GeV,
with $\tan\beta =5$ and $\mu >0$.}
\bigskip
\begin{tabular}{lcccc}
\hline
parameter & $\Phi({\bf 1})$ & $\Phi({\bf 24})$ & $\Phi({\bf 75})$ & 
$\Phi({\bf 200})$ \\
\hline
$m_{\tg}$     & 394.9 & 397.7 & 409.3 & 404.0 \\
$m_{\tu_L}$   & 356.8 & 372.0 & 457.9 & 406.2 \\
$m_{\tst_1}$  & 243.7 & 283.8 & 255.2 & 295.0 \\
$m_{\tb_1}$   & 328.8 & 342.0 & 356.5 & 340.2 \\
$m_{\tell_L}$ & 154.4 & 191.1 & 360.1 & 372.7 \\
$m_{\tell_R}$ & 123.2 & 112.3 & 310.4 & 589.8 \\
$m_{\ttau_1}$ & 120.9 & 111.6 & 309.8 & 372.1 \\
$m_{\tw_1}$   & 95.6  & 147.0 & 93.2  & 156.3 \\
$m_{\tz_2}$   & 99.7  & 142.5 & 106.0 & 202.5 \\
$m_{\tz_1}$   & 53.2  &  33.1 & 92.6  & 151.9 \\
$m_h$         & 103.4 &  99.8 & 104.5 & 106.7 \\
$m_A$         & 257.2 & 249.6 & 372.9 & 421.3 \\
$\mu$         & 215.2 & 173.0 & 104.7 & 197.2 \\
\hline
\label{tnusug}
\end{tabular}
\end{center}
\end{table}
\begin{table}
\begin{center}
\caption{Model parameters and weak scale sparticle masses in GeV 
for mSUGRA and for the MSSM+right-handed neutrino model. For each
case, we take $m_0=200$ GeV, $m_{1/2}=200$ GeV, $A_0=0$, $\tan\beta =40$
and $\mu >0$.}
\bigskip
\begin{tabular}{lcc}
\hline
parameter & $mSUGRA$ & $MSSM+RHN$ \\
\hline
$m_{\nu_\tau}$ & 0 & $10^{-9}$ \\
$M_N$ & --- & $10^{13}$ \\
$m_{\tnu_{\tau R}}$ & --- & 200 \\
$A_{\nu}$ & --- & 0 \\
$m_{\tg}$ & 511.5 & 511.5 \\
$m_{\tu_L}$ & 485.1 & 485.1 \\
$m_{\tst_1}$ & 343.1 & 344.2 \\
$m_{\tb_1}$ & 386.6 & 386.4 \\
$m_{\tell_L}$ & 250.4 & 250.4 \\
$m_{\tell_R}$ & 218.9 & 218.9 \\
$m_{\ttau_1}$ & 144.2 & 140.5 \\
$m_{\ttau_2}$ & 257.1 & 252.6 \\
$m_{\tnu_\tau}$ & 220.1 & 211.6 \\
$m_{\tw_1}$ & 146.9 & 147.6  \\
$m_{\tz_2}$ & 147.5 & 148.2  \\
$m_{\tz_1}$ & 79.9 &  80.0 \\
$m_h$ & 111.7 & 111.7 \\
$m_A$ & 243.3 & 249.4 \\
$\mu$ & 263.2 & 267.6 \\
$A_\tau$ & -66.6 & -66.5 \\
$A_t$ & -383.3 & -383.3 \\
\hline
\label{trhn}
\end{tabular}
\end{center}
\end{table}
\begin{table}
\begin{center}
\caption{Model parameters and weak scale sparticle masses for mSUGRA
and Yukawa unified $SO(10)$ with $D$-terms and GUT symmetry breaking
at $M_{GUT}$. Note the mSUGRA case has a somewhat smaller value
of $\tan\beta$ than $SO(10)$, since no mSUGRA solution could be
obtained with $\tan\beta =48.6$.}
\bigskip
\begin{tabular}{lcc}
\hline
parameter & mSUGRA & $SO(10)$ \\
\hline

$m_{16}$ & 1022.0 & 1022.0 \\
$m_{10}$ & 1022.0 & 1315.0 \\
$M_D$    & 0.0    & 329.8  \\
$m_{1/2}$& 232.0  & 232.0  \\
$A_0$    & -1350.0& -1350.0\\
$\tan\beta$ & 45.0& 48.6   \\
$m_{\tg}$ & 639.0 & 631.5  \\
$m_{\tu_L}$ & 1130.6& 1178.5 \\
$m_{\td_R}$ & 1121.7& 970.1 \\
$m_{\tst_1}$& 553.0 & 512.3 \\
$m_{\tb_1}$ & 657.1 & 187.1 \\
$m_{\tell_L}$ & 1035.8 & 857.8 \\
$m_{\tell_R}$ & 1026.8 & 1088.9 \\
$m_{\tnu_{e}}$& 1032.8 & 854.1 \\
$m_{\ttau_1}$ & 725.4  & 623.6 \\
$m_{\tnu_{\tau}}$ & 897.8 & 619.5 \\
$m_{\tw_1}$ & 193.5 & 122.9 \\
$m_{\tz_2}$ & 193.3 & 131.6 \\
$m_{\tz_1}$ & 97.4  & 84.0  \\
$m_h$ & 88.5 & 118.8 \\
$m_A$ & 90.4 & 479.9 \\
$m_{H^+}$ & 131.2 & 490.2 \\
$\mu$ & -547.8 & -150.5 \\
$\langle\ttau_1 |\ttau_L\rangle$ &
0.14 & 0.99 \\
\hline
\label{tso10_1}
\end{tabular}
\end{center}
\end{table}
\begin{table}
\begin{center}
\caption{Model parameters and weak scale sparticle masses for 
Yukawa unified $SO(10)$ with $D$-terms. The first model has
universality of matter scalars at $M_{GUT}$, while the second
has universality at $M_P$. For both cases, we take $m_{16}=629.8$ GeV and
$m_{10}=836.2$ GeV. At $M_{GUT}$, for both cases, we take
$m_{1/2}=348.8$ GeV, $A=-186.5$ GeV. We also take $\mu <0$, 
$\tan\beta =52.1$ and $M_D=135.6$, where $D$-terms are imposed
at $M_{GUT}$.}
\bigskip
\begin{tabular}{lcc}
\hline
parameter & $M_{GUT}$ Unification & $M_P$ Unification \\
\hline

$m_{Q_1}$ & 644.2 & 677.2 \\
$m_{L_1}$ & 584.4 & 621.0 \\
$m_{Q_3}$ & 644.2 & 603.5 \\
$m_{L_3}$ & 584.4 & 539.2 \\
$m_{H_d}$ & 857.9 & 833.9 \\
$m_{H_u}$ & 813.9 & 788.6 \\
$m_{\tg}$ & 813.9 & 838.1 \\
$m_{\tu_L}$ & 974.4 & 969.9 \\
$m_{\td_R}$ & 910.8 & 914.3 \\
$m_{\tst_1}$& 618.7 & 586.2 \\
$m_{\tb_1}$ & 636.8 & 600.1 \\
$m_{\tell_L}$ & 634.6 & 660.6 \\
$m_{\tell_R}$ & 662.5 & 692.7 \\
$m_{\tnu_{e}}$& 629.5 & 655.8 \\
$m_{\ttau_1}$ & 427.8 & 397.8 \\
$m_{\tnu_{\tau}}$ & 519.1 & 474.0 \\
$m_{\tw_1}$ & 106.3 & 130.0 \\
$m_{\tz_2}$ & 126.1 & 153.9 \\
$m_{\tz_1}$ & 87.5  & 105.2  \\
$m_h$ & 93.7 & 104.7 \\
$m_A$ & 93.9 & 105.2 \\
$m_{H^+}$ & 137.1 & 144.9 \\
$\mu$ & -113.9 & -142.4 \\
\hline
\label{msoten}
\end{tabular}
\end{center}
\end{table}
\begin{table}
\begin{center}
\caption{Model parameters and weak scale sparticle masses for mSUGRA
model and general $SO(10)$. The first model has universality of matter
scalars at $M_{GUT}$, while the second has universality at $M_P$. For
both cases, we take $m_{0}=150$ GeV, $m_{1/2}=200$ GeV, $A_0=0$,
$\tan\beta =35$ and $\mu <0$, and list the values of the SSB masses at
the $GUT$ scale. This yields $f_t= 0.534$ and $f_b =f_\tau =0.271$, with
$g(M_{GUT})=0.717$ as in Table 1.  }
\bigskip
\begin{tabular}{lcc}
\hline
parameter & mSUGRA & $SO(10)$ \\
\hline

$m_{16_1}$ & 150.0 & 195.0 \\
$m_{16_3}$ & 150.0 & 177.0 \\
$m_{H_d}$ & 150.0 & 183.9 \\
$m_{H_u}$ & 150.0 & 175.1 \\
$m_{H_{12}}$ & --- & -46.6 \\
$A_t(M_{GUT})$ & 0.0 & -117.8 \\
$A_b(M_{GUT})$ & 0.0 & -120.0 \\
$m_{\tg}$ & 512.0 & 492.9 \\
$m_{\tu_L}$ & 468.0 & 468.1 \\
$m_{\td_R}$ & 454.7 & 457.2 \\
$m_{\tst_1}$& 335.6 & 319.1 \\
$m_{\tb_1}$ & 375.4 & 360.1 \\
$m_{\tell_L}$ & 212.8 & 241.3 \\
$m_{\tell_R}$ & 174.5 & 213.2 \\
$m_{\tnu_{e}}$& 197.2 & 227.7 \\
$m_{\ttau_1}$ & 124.3 & 139.6 \\
$m_{\tnu_{\tau}}$ & 189.3 & 201.4 \\
$m_{\tw_1}$ & 150.1 & 142.5 \\
$m_{\tz_2}$ & 150.3 & 142.4 \\
$m_{\tz_1}$ & 80.8  & 77.8 \\
$m_h$ & 111.0 & 111.4 \\
$m_A$ & 210.4 & 218.7 \\
$m_{H^+}$ & 227.9 & 235.6 \\
$\mu$ & -263.8 & -285.2 \\
\hline
\label{gsoten}
\end{tabular}
\end{center}
\end{table}
\begin{table}
\begin{center}
\caption{Model parameters and weak scale sparticle masses in GeV for 
an anomaly-mediated SUSY breaking (AMSB) case study.}
\bigskip
\begin{tabular}{l|cc}
\hline
parameter & $AMSB$(200) & $AMSB$(500) \\
\hline
$m_0$ & 200 &500 \\
$m_{3/2}$ & 35,000 & 35,000\\
$\tan\beta$ & 5 & 5 \\
$\mu$ & $>0$ & $>0$ \\
$m_{\tg}$ & 816 & 825 \\
$m_{\tu_L}$ & 754 & 872 \\
$m_{\tst_1}$ & 512 & 588 \\
$m_{\tb_1}$ & 666 & 758 \\
$m_{\tell_L}$ & 156 & 484 \\
$m_{\tell_R}$ & 154 & 483  \\
$m_{\ttau_1}$ & 132 & 476  \\
$m_{\ttau_2}$ & 173 & 489 \\
$m_{\tw_1}$ & 99.5 & 99.1  \\
$m_{\tz_2}$ & 321 & 321 \\
$m_{\tw_1}-m_{\tz_1}$ & 0.171 & 0.172 \\
$m_h$ & 114 & 113 \\
$m_A$ & 654 & 806 \\
$\mu$ & 632 & 635 \\
$\theta_{\tau}$ & 0.82 & 0.86 \\
$\theta_b$ & 0.11 & 0.074 \\
\hline
\label{tamsb}
\end{tabular}
\end{center}
\end{table}
\begin{table}
\begin{center}
\caption{Input and output parameters for the Minimal Gaugino Mediation
model case study described in the text.}
\bigskip
\begin{tabular}{lcc}
\hline
parameter & scale & value\\
\hline
$m_0$ & $M_c$ & 0 \\
$A_0$ & $M_c$ & 0 \\
$m_{1/2}$ & $M_{GUT}$ & 300 \\
$g_5$ & $M_{GUT}$ & 0.717  \\
$f_t$ & $M_{GUT}$ & 0.534  \\
$f_b=f_\tau$ & $M_{GUT}$ & 0.271 \\
$\lambda$ & $M_{GUT}$ & 1 \\
$\lambda'$ & $M_{GUT}$ & 0.1 \\
$\tan\beta$ & $M_{Weak}$ & 35 \\
$\mu $ & $M_{Weak}$ & $<0$ \\
\hline
$m_{\tg}$ & $M_{Weak}$ & 737.2 \\
$m_{\tu_L}$ & $M_{Weak}$ & 668.5 \\
$m_{\td_R}$ & $M_{Weak}$  & 633.1 \\
$m_{\tst_1}$ & $M_{Weak}$ & 482.8 \\
$m_{\tb_1}$ & $M_{Weak}$ & 541.5 \\
$m_{\tell_L}$ & $M_{Weak}$ & 258.6 \\
$m_{\tell_R}$ & $M_{Weak}$ & 210.0 \\
$m_{\ttau_1}$ & $M_{Weak}$ & 143.3 \\
$m_{\tw_1}$ & $M_{Weak}$ & 240.2  \\
$m_{\tz_2}$ & $M_{Weak}$ & 240.0  \\
$m_{\tz_1}$ & $M_{Weak}$ &  124.8 \\
$m_h$ & $M_{Weak}$ & 115.6 \\
$m_A$ & $M_{Weak}$ & 311.2.4 \\
$\mu$ & $M_{Weak}$ & -411.5 \\
\hline
\label{mgmtable}
\end{tabular}
\end{center}
\end{table}
\begin{table}
\begin{center}
\caption{Model parameters and weak scale sparticle masses in GeV for
$O$-$I$ and $O$-$II$ models discussed in the text. For the $O$-$I$ model, we
take $\tan\beta=4$, $n_{H_u}=-3$, $\sin\theta=0.85$
and other parameters as in Fig.~\ref{oimass}. For the $O$-$II$ model, we
fix $\tan\beta=4$, $\delta_{GS}=-5$ and illustrate the results for the
three cases discussed in the text. For both models we take $\cos\theta>0$.
We have fixed the string scale to be
$4\times 10^{17}$~GeV.}
\bigskip
\begin{tabular}{l|cccc}
\hline
Mass & $O$-$I$ & $O$-$II$ ($\sin\theta=0.85)$ &$O$-$II$ ($\sin\theta=0)$ 
&$O$-$II$ ($\sin\theta=0.005)$ \\
\hline
$m_{\tg}$ & 698 & 773 & 360 & 1693 \\
$m_{\tu_L}$ & 633 & 701 & 4241 & 4436 \\
$m_{\tu_R}$ & 602 & 677 & 4239 & 4409  \\
$m_{\td_R}$ & 607 & 673 & 4237 & 4400  \\
$m_{\tst_1}$& 442 & 523 & 2468 & 2637 \\
$m_{\tst_2}$ & 651 & 708& 3478 & 3672 \\
$m_{\tb_1}$ & 583 & 650 & 3475 & 3669 \\
$m_{\tb_2}$ & 608 & 674 & 4232 & 4394 \\
$m_{\tell_L}$ & 223 & 292 & 4239& 4289\\
$m_{\tell_R}$ & 132 & 223 & 4240& 4272\\
$m_{\tw_1}$ & 202 & 235 & 254 & 715\\
$m_{\tz_1}$ & 99 & 124 & 249 & 490   \\
$m_{\tz_2}$ & 202 & 236 & 255 & 715 \\
$m_h$ & 109 & 108 & 116 & 117 \\
$m_A$ & 497 &516  & 4506 & 4634\\
$\mu$ & 422 & 416 & 1114 & 1379\\
\hline
\label{o-scenarios}
\end{tabular}
\end{center}
\end{table}

\newpage
%
%%%%%%%%%%%%%%%%%%%%%%%%%%%%%%%%%%%%%%%%%%%%%%%%%%%%%%%%%%%%%%%%%%%%%%%%%%%%%

%%%%%%%%%%%%%%%%%%%%%% FIGURE CAPTIONS %%%%%%%%%%%%%%%%%%%%%%%%%%%%%%%%%%%%%%
%
\begin{figure}
\dofig{4.5in}{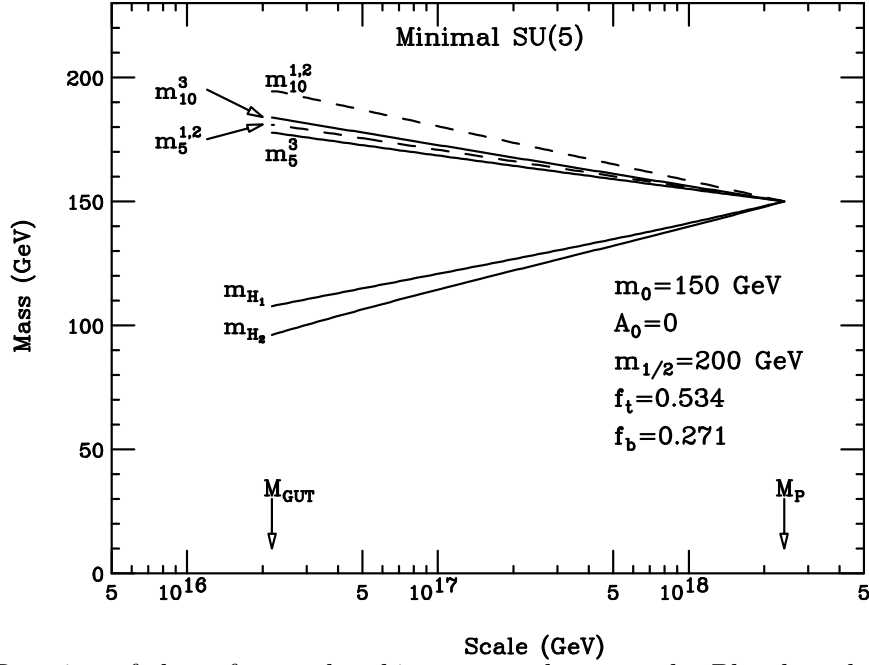}
\caption[]{
Running of the soft susy breaking masses between the Planck scale 
and the GUT scale in the minimal $SU(5)$ model for $\tan\beta =35$.
At the GUT scale we
have taken $\lambda=1$ and $\lambda'=0.1$ for the Higgs couplings, and 
$\alpha_{GUT}=0.041$ for the unified gauge coupling.}
\label{mpmg1}
\end{figure}
\begin{figure}
\dofig{4.5in}{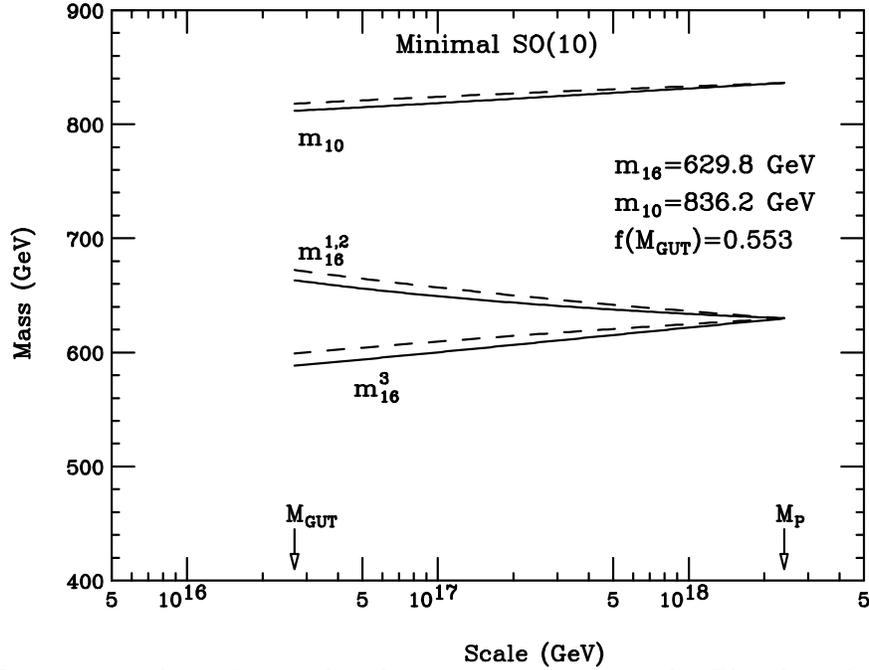}
\caption[]{
Running of the soft susy breaking masses between the Planck scale 
and the GUT scale in the {\it minimal} $SO(10)$ model.}
\label{so10_min}
\end{figure}
\begin{figure}
\dofig{5in}{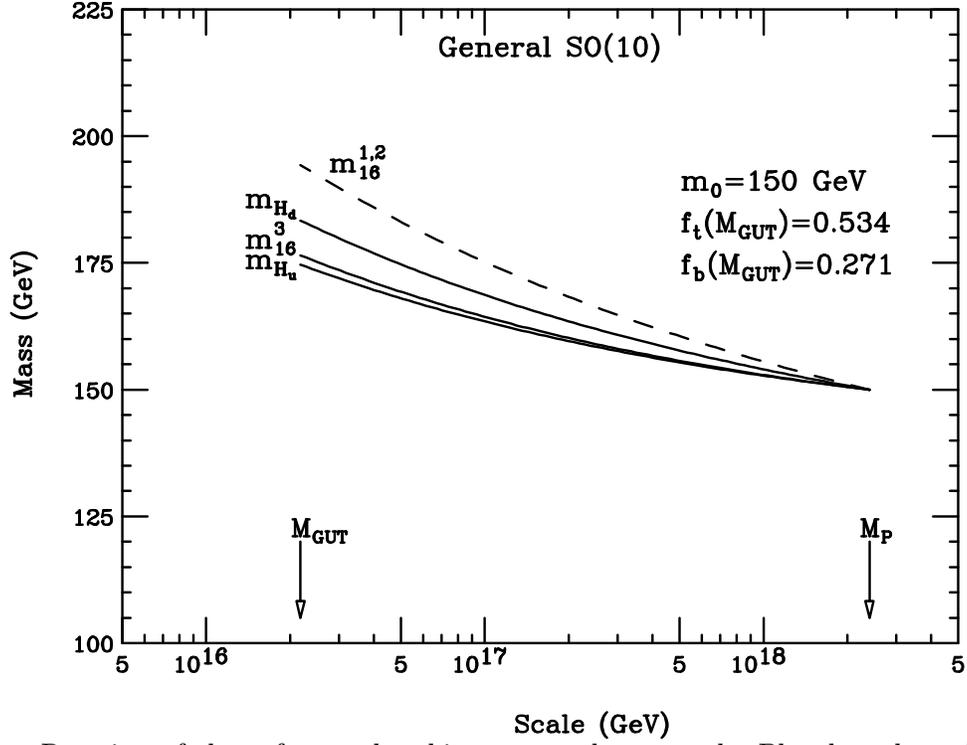}
\caption[]{
Running of the soft susy breaking masses between the Planck scale 
and the GUT scale in the general $SO(10)$ model. The GUT scale
Yukawa couplings here are the same as in the $SU(5)$ case.}
\label{so10_gen}
\end{figure}
\begin{figure}
\dofig{4.5in}{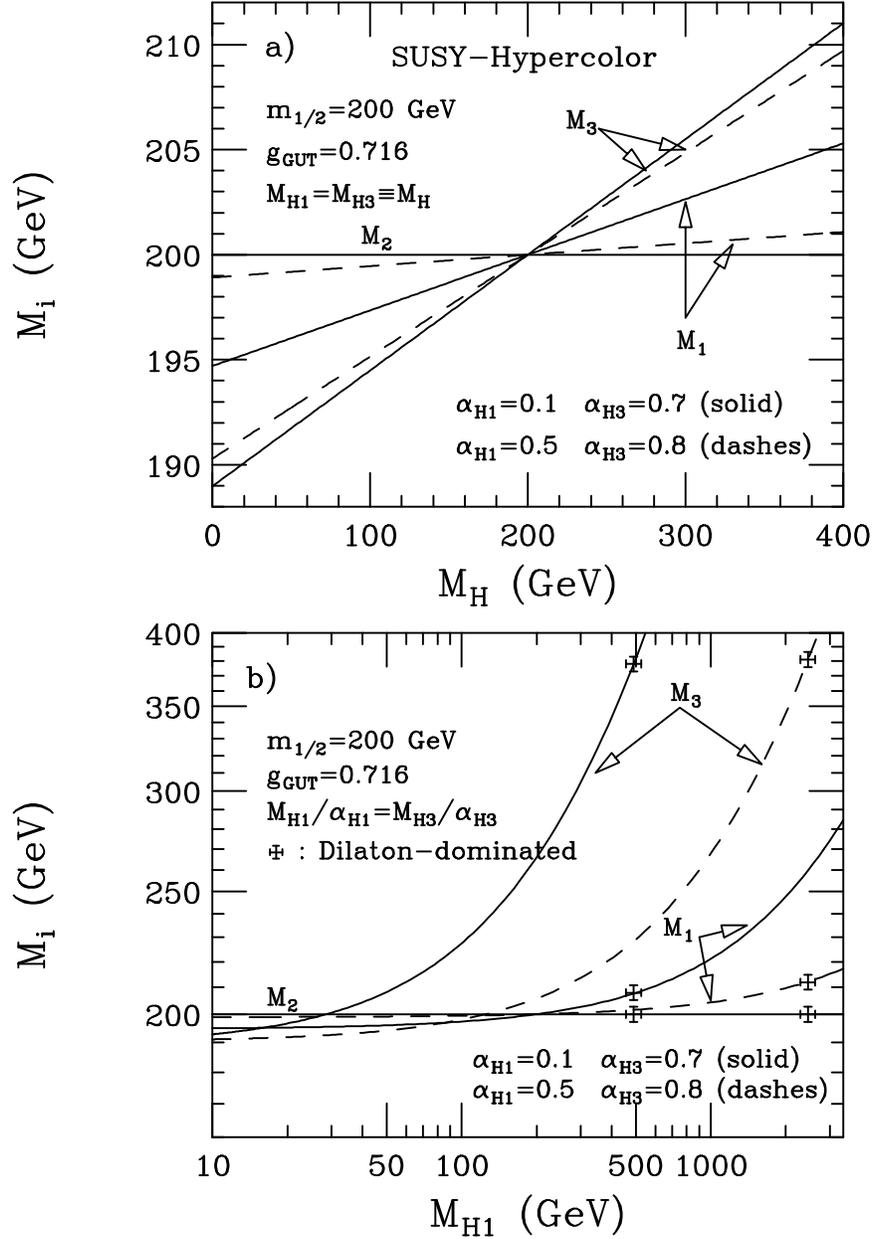}
\caption[]{
Non-universal gaugino masses $M_i$, $i=1,2,3$, in a supersymmetric missing
partner model with hypercolor, as a function of the common gaugino mass
$M_{H1}=M_{H3}\equiv M_H$ in the hypercolor sector. Two values of the
hypercolor gauge couplings are used.
}
\label{hyperc}
\end{figure}
\begin{figure}
\dofig{7in}{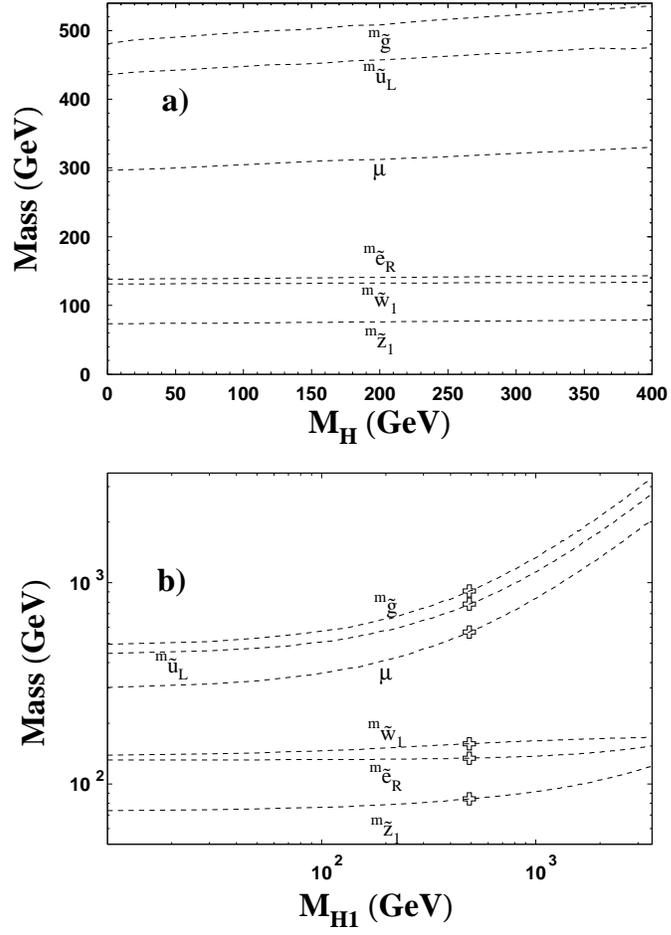}
\caption[]{
Various sparticle masses in the hypercolor model.}
\label{hyper1}
\end{figure}
\begin{figure}
\dofig{4.5in}{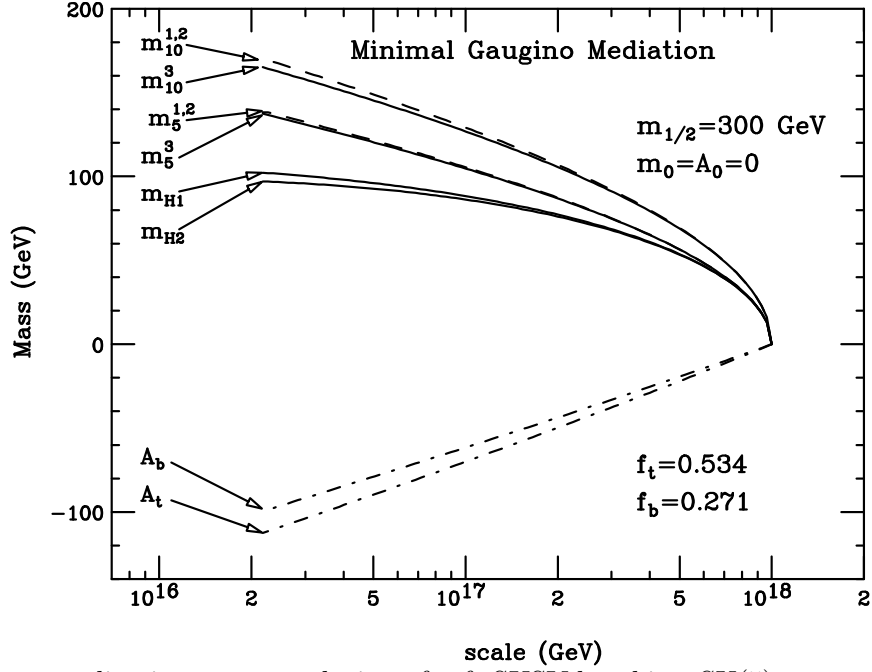}
\caption[]{
Renormalization group evolution of soft SUSY breaking $SU(5)$ masses
versus scale in the minimal gaugino mediation model. We take $\tan\beta =35$
and $\mu <0$ to achieve $b-\tau$ Yukawa coupling unification.
}
\label{mgmrun}
\end{figure}
\begin{figure}
\dofig{4.5in}{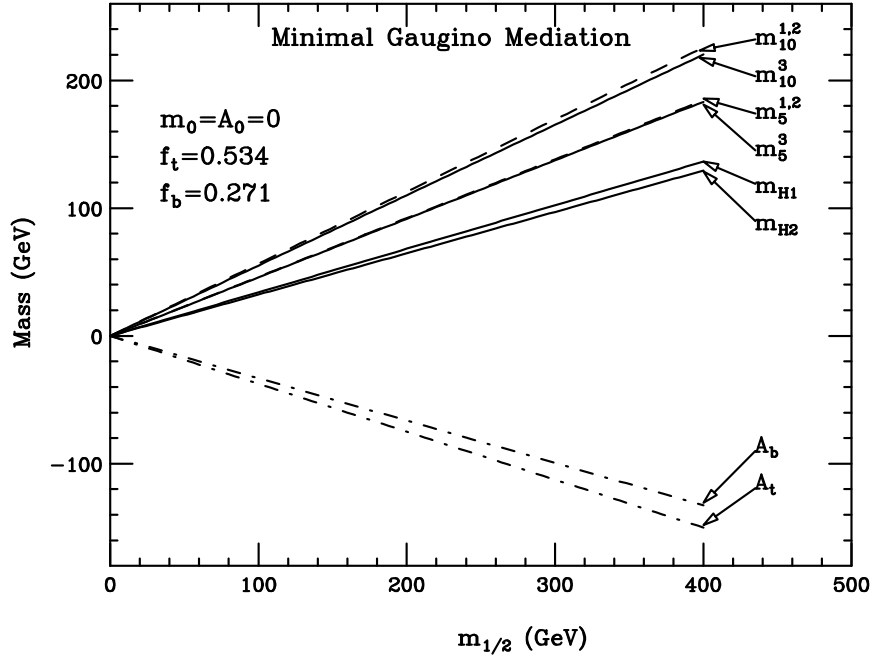}
\caption[]{
$GUT$ scale values of $SU(5)$ SSB masses in the minimal gaugino mediation
model. We take $\tan\beta =35$ and $\mu <0$ to achieve $b-\tau$ Yukawa 
coupling unification. We take the compactification scale 
$M_c =1\times 10^{18}$ GeV. Models with $m_{1/2}< 275$ GeV lead to
a breakdown in REWSB or a charged LSP.}
\label{mgmmhalf}
\end{figure}
%
%
%\begin{figure}
%\dofig{4.5in}{cylt.ps}
%\caption[]{
%Universal gaugino ($m_{1/2}$) and scalar ($m_0$) masses as a function
%of $\sin\theta$ in Calabi--Yau compactifications in the large--T limit
%for two values of the cosmological constant.
%}
%\label{cylt}
%\end{figure}
%
%
%\begin{figure}
%\dofig{4.5in}{cygen.ps}
%\caption[]{
%Non--universal scalar masses $m_1$ and $m_2$ as a function of $\sin\alpha$
%in general Calabi--Yau compactification with two moduli. The angle $\alpha$
%defines the direction of the vev in the moduli space. Two values of the
%goldstino angle $\theta$ are displayed.
%}
%\label{cygen}
%\end{figure}
%
%
\begin{figure}
\dofig{4.5in}{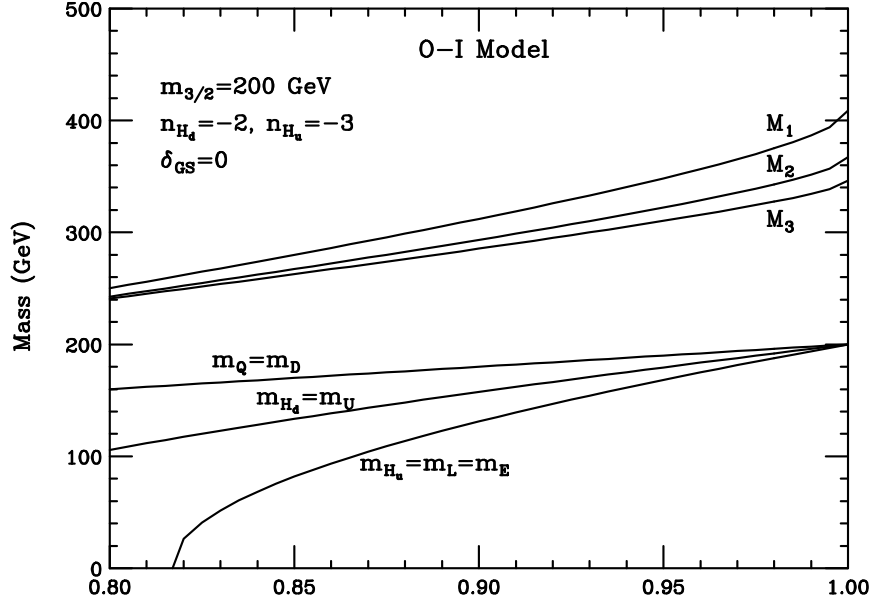}
\caption[]{
Soft SUSY breaking masses in the $O$-$I$ superstring model, versus 
$\sin\theta$, for $m_{3/2}=200$ GeV, $n_{H_D}+n_{H_u}=-5$
and $\delta_{GS}=0$. The $O$-$I$ model assumes $n_{Q}=n_{D}=-1$, 
$n_{U}=-2$, and $n_{L}=n_{E}=-3$.
}
\label{oimass}
\end{figure}
\begin{figure}
\dofig{4.5in}{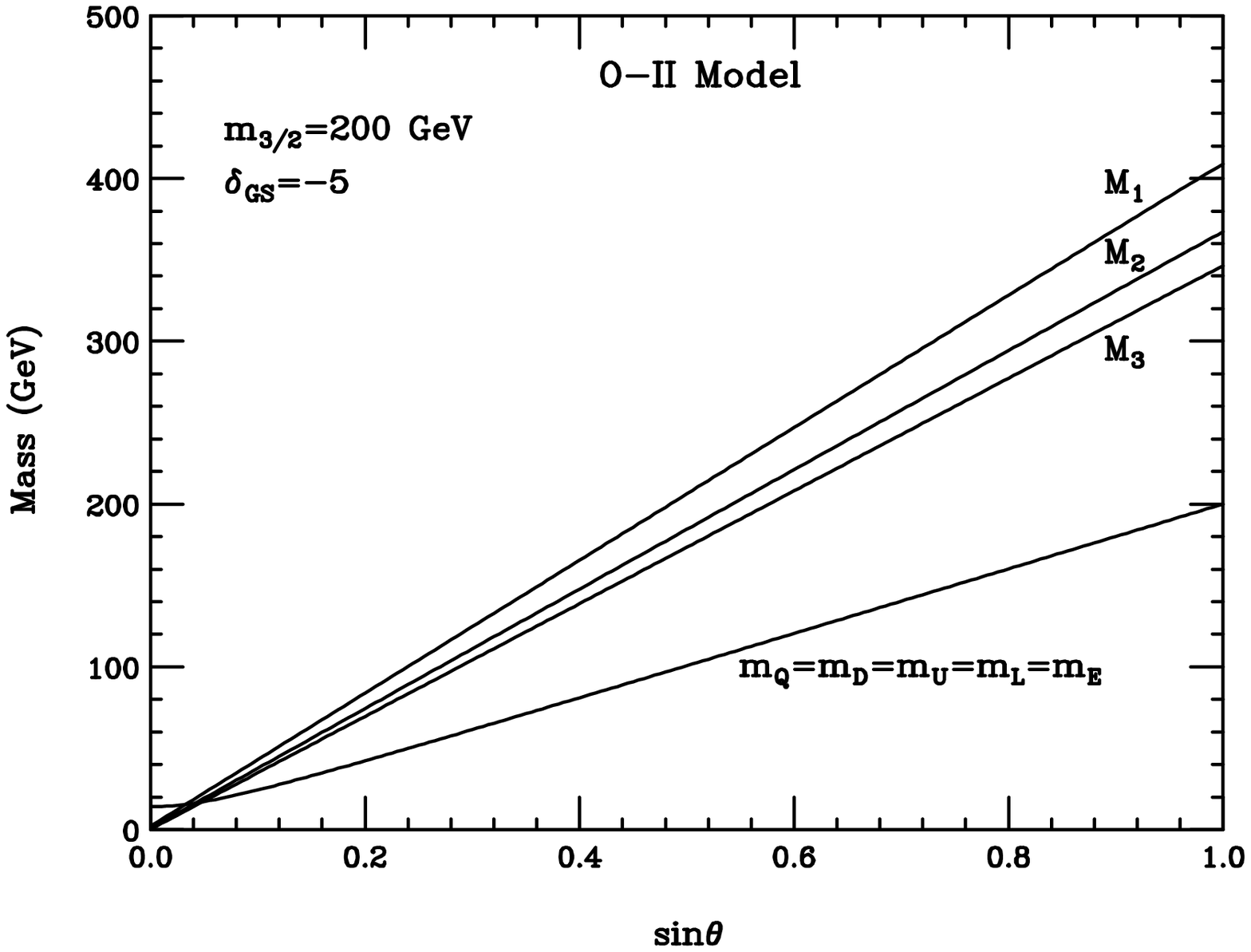}
\caption[]{
Soft SUSY breaking masses in the $O$-$II$ superstring model, versus
$\sin\theta$, for $m_{3/2}=200$ GeV and $\delta_{GS}=-5$. The $O$-$II$ 
model assumes $n_{Q}=n_{D}=n_{U}=n_{L}=n_{E}=-1$.
}
\label{oiimass}
\end{figure}
\begin{figure}
\dofig{4.5in}{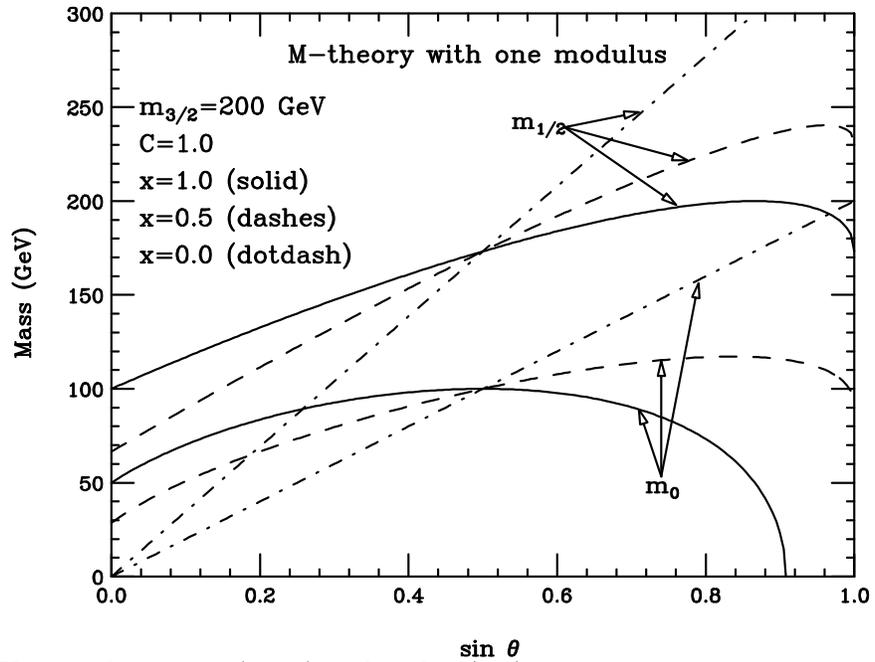}
\caption[]{
Universal gaugino ($m_{1/2}$) and scalar ($m_0$) masses as a function
of $\sin\theta$ in M--theory with one modulus for three values of the parameter
$x$ and zero cosmological constant.
}
\label{m1mod}
\end{figure}

\end{document}